\documentclass{jpsj3}
\usepackage{graphicx}
\usepackage{dcolumn}
\usepackage{bm}
\usepackage{color}
\title{%
  Weak Localization Correction to Linear 
  Absorption in Conventional Superconductors
}
\author{%
  Takanobu Jujo\thanks{E-mail address: jujo@ms.aist-nara.ac.jp}
}
\inst{%
  Division of Materials Science,
  Graduate School of Science and Technology,
  Nara Institute of Science and Technology,
  Ikoma, Nara 630-0101, Japan
}
\abst{%
  The weak localization effect
  on a linear absorption spectrum
  is investigated for disordered s-wave superconductors.
  The vertex correction is incorporated into the response function
  in a way that is consistent with
  the weak localization correction to a one-particle spectrum.
  The effect of interactions between electrons enhanced by their
  diffusive motion makes a major contribution to the correction term.
  Numerical calculations show that
  the weak localization effect suppresses 
  the absorption 
  by the excitation across the gap
  to a greater extent than that by thermal excitation,
  which is a similar tendency to the results of experiments.
  The ratio of
  the linear absorption with vertex corrections to that without
  the corrections
  shows a large variation
  at   the frequency that is around  twice the energy of
  the superconducting gap.
  This behavior represents a relationship between 
  the weak localization effect on the linear absorption and 
  that on the one-particle spectrum.
}

\begin{document}
\setlength{\textwidth}{504pt}
\setlength{\columnsep}{14pt}
\hoffset-23.5pt
\maketitle

\section{Introduction}

The weak localization effect has so far been
studied mainly in the normal state.
In the case of the noninteracting electron systems,
a correction term in conductivity
arises from the coherent backscattering by impurities.~\cite{gorkov}
The motion of electrons becomes diffusive
in the presence of the impurity scattering,
and electron--electron interactions
such as the screened Coulomb interaction are
modified.~\cite{schmidA,AASSC,AAJETP}
This effect
also gives a weak localization correction
to the dc and ac conductivities.~\cite{AAJETP,fukuyama,AALee}

Through the same effect, the superconducting transition temperature
$T_c$
is lowered because the interactions between electrons
change owing to the impurity scattering.~\cite{ovchinnikov,maekawa,takagi}
In the case of the superconducting state, there have been few studies
on weak localization effect other than the correction to $T_c$.
For example, superfluid density is calculated
considering only the effect of backscattering by impurities.~\cite{smith}
As for the conductivity, the influence of superconducting
fluctuation becomes strong
near the transition temperature.~\cite{aslamazovL,maki,thompson}
This fluctuation effect on ac conductivity
has been investigated
in the normal state,~\cite{schmidt,aslamasovV,federici,petkovic}
but the effect of the Coulomb interaction has not been taken into account
in these studies.

In the presence of the superconducting gap ($\Delta$)
(when the temperature is below the transition temperature),
the real part of 
ac conductivity takes finite values at a frequency ($\omega$)
lower than twice the superconducting gap
owing to thermally excited quasiparticles.
In the case of $\omega>2\Delta$, there exists a finite absorption 
even at absolute zero owing to
the excitation across the gap.~\cite{mattis,abrikosov}
This behavior of ac conductivity is described by 
the Mattis--Bardeen (MB) formula.

In recent years, deviations from the MB formula 
have been observed experimentally in
strongly disordered systems.~\cite{cheng,simmendinger,pracht}
These studies show that 
the absorption for $\omega<2\Delta$ is large
and the spectrum at the gap edge ($\omega\simeq 2\Delta$)
becomes blurred
as compared with that obtained on the basis of the MB theory.
This phenomenon has been interpreted with
several ideas, such as
the pair-breaking effect by nonuniformity,~\cite{larkin}
collective excitation modes,~\cite{pracht2017}
or the existence of a normal (Drude) component.~\cite{simmendinger}
These systems are considered to be situated near
the superconductor--insulator transition and
have inhomogeneities.
It is yet unknown whether
these systems reflect the weak localization effect
partly because there is no theory about this correction effect in
the superconducting state.

In this study, we show how
the weak localization effect appears in the ac conductivity
of superconductors in a homogeneously disordered system.
The conductivity including vertex corrections is calculated
in a three-dimensional system, in which
the expansion parameter is $1/(k_Fl)^2$ where
$k_F$ is the Fermi wave number and $l=v_F\tau$
is the mean free path ($v_F$ and $\tau$
being the Fermi velocity and the relaxation time, respectively).
We derive vertex corrections from
the functional derivative of self-energy.
The latter
gives a weak localization correction to a one-particle spectrum
in the superconducting state.~\cite{jujo2019}
According to the calculated results, it is found that the weak
localization correction is larger for $\omega>2\Delta$
is larger than for $\omega<2\Delta$.
Taking
the ratio of the correction term to the MB conductivity
clarifies how the correction effect on
the linear absorption is related to
that on the one-particle spectrum.

The structure of this paper is as follows.
Section 2 gives a formulation for calculating response functions
including the impurity scattering and interactions between electrons.
In Sect. 3, a formula for the conductivity including vertex corrections
that give a weak localization effect is obtained.
Section 4 gives the results of numerical calculations
on the basis of obtained expressions for the conductivity.

\section{Formulation}

We calculate the ac conductivity $\sigma_q$
by Keldysh's method~\cite{keldysh}
with the use of the functional integral.~\cite{kamenev}
The absorption spectrum is given by the
real part of the ac conductivity, and it is written as 
\begin{equation}
  {\rm Re}\sigma_q=-\frac{{\rm Im}K_q}{\omega}.
\end{equation}
``${\rm Re}$'' and ``${\rm Im}$'' indicate
the real and imaginary parts, respectively.
$q=(\mib q,\omega)$ with
$\mib q$ being the wave number vector
in three dimensions. We consider only the uniform case
$(\mib q=\mib 0)$ and finite frequencies ($\omega\ne 0$),
and set $\hbar=c=1$ in this paper.
$K_q$ is a response function and defined with 
the current density $J_q$ as 
\begin{equation}
  J_q=-K_q A_q,
  \label{eq:JKA}
\end{equation}
where $A_q$ is the vector potential.
We consider an isotropic system and omit
indices of vectors.
The current density is derived from the functional derivative
of the action ($S$) by the vector potential:
\begin{equation}
  J_q=\left.\frac{-i}{\sqrt{2}}\frac{\delta
    {\rm ln}\langle{\rm e}^{i S}\rangle_{e,p,i}}
  {\delta A_{-q}^{qu}}\right|_{A^{qu}\to 0}.
  \label{eq:JdSdA}
\end{equation}
Here,
$A^{qu}_q$ is the Fourier transform of
$A^{qu}_{\mib q,t}:=(A^+_{\mib q,t}-A^-_{\mib q,t})/\sqrt{2}$
with $A^{+(-)}_{\mib q,t}$ being the vector potential
in the forward (backward) direction in time.~\cite{kamenev}
The vector potential in Eq. (\ref{eq:JKA})
is given by the Fourier transform of 
$(A^+_{\mib q,t}+A^-_{\mib q,t})/2$
$(=:A^{cl}_{\mib q,t}/\sqrt{2})$.
We consider the following action:
\begin{equation}
  \begin{split}
  S=&\int_{\cal C} d t\{
  \sum_{\mib k,\sigma}\bar{\psi}_{\mib k,\sigma,t}
  (i\partial_t-\xi_{\mib k})\psi_{\mib k,\sigma,t}
    -
  (2N^3)^{-1}\sum_{\mib k,\mib k',\mib q,\sigma,\sigma'}v_{\mib q}^C
  \bar{\psi}_{\mib k,\sigma,t}\psi_{\mib k+\mib q,\sigma,t}
  \bar{\psi}_{\mib k',\sigma',t}\psi_{\mib k'-\mib q,\sigma',t}\\
  &  +
 \sum_{\mib q}\bar{b}_{\mib q,t}
  (i\partial_t-\omega_{\mib q})b_{\mib q,t}
+
  (N^3)^{-1/2}\sum_{\mib k,\mib q,\sigma}\bar{\psi}_{\mib k+\mib q,\sigma,t}
  [e{\mib A}_{\mib q,t}\cdot{\mib v}_{\mib k+\mib q/2}
   -g_{ph} \phi_{\mib q,t}
   -   u_{\mib q}]
  \psi_{\mib k,\sigma,t}
  \}.
   \end{split}
\end{equation}
(The integration $\int_{\cal C} d t$ is taken over the forward
and backward time contour.~\cite{kamenev} 
$\xi_{\mib k}$ and $\omega_{\mib q}$ are
dispersions of electrons and phonons, respectively.
The terms including $v^C_{\mib q}=4\pi e^2/{\mib q}^2$,
$g_{ph}$, and $u_{\mib q}$ describe
the Coulomb interaction between electrons, 
the electron--phonon coupling, and the impurity scattering,
respectively.
We take into account only the first order of
the external field ${\mib A}_q$ 
because we consider the linear absorption.
$N^3$ is the number of sites,
and ${\mib v}_{\mib k}=\partial \xi_{\mib k}/\partial {\mib k}$.
$\phi_{\mib q,t}:=b_{\mib q,t}+\bar{b}_{-\mib q,t}$.)
$\langle\cdot\rangle_{e,p,i}$ means the integrations
over the degrees of freedom of electrons and phonons
and averaging over the impurities.

Firstly, we integrate out
the electronic degrees of freedom
($\langle{\rm e}^{i S}\rangle_{e,p,i}
=\langle{\rm e}^{i S'}\rangle_{\varphi,p,i}$).
Here, 
\begin{equation}
  \begin{split}
    i S'=&\frac{1}{2}\sum_{\mib q}\int d\omega
    \sum_{l,l'=cl,qu}
  \left[
  \frac{-i\omega_E}{2}  \phi^l_{-q}
m^{ph}_{l,l'}
  \phi^{l'}_{q}
  +
  \frac{-i e^2}{v^C_{\mib q}}    \varphi^{l}_{-q}
m^{v}_{l,l'}
  \varphi^{l'}_q
  \right]
  +\sum_{n=1}^{\infty}\frac{(-1)^{n-1}}{n}{\rm tr}(G V)^n.
  \end{split}
  \label{eq:actionSd}
\end{equation}
The first term describes the phonon degrees of freedom,
and the second term is obtained from the Hubbard--Stratonovich
transformation of the Coulomb interaction
between electrons.~\cite{HStrans}
$cl$ $(qu)$ means a sum (difference)
of the forward and backward paths in time
divided by $\sqrt{2}$
(this definition is different from that in Ref. 27 
  by a factor of $\sqrt{2}$).
$m^{ph}_{l,l'}=m^{v}_{l,l'}=1$ for $(l,l')=(cl,qu)$ and $(qu,cl)$,
and $m^{ph}_{l,l'}=m^{v}_{l,l'}=0$ for $(l,l')=(cl,cl)$ and $(qu,qu)$.
In Eq. (\ref{eq:actionSd}), we adapted approximations that
the dispersion of phonons takes a constant value ($\omega_E$)
and the electron--phonon interaction is weak coupling
(the effect of retardation is omitted: $\omega-\omega_E\simeq -\omega_E$).
$G$ is Green's function of electrons
and $V$ describes interaction effects.
$V$ and $G$ are written as $4\times4$ matrices
(the product of Keldysh and Nambu spaces).
${\rm tr}[\;\cdot\;]$ indicates the trace over $4\times 4$
matrices and includes the summation over wave numbers
and integration over frequencies.
$\hat{\;\cdot\;}$ and ${\rm Tr}[\;\cdot\;]$
given below
indicate $2\times 2$ Nambu matrices and the trace over
these matrices, respectively.
\begin{equation}
  G=\begin{pmatrix}
  \hat{G}^+_k & \hat{G}^K_k \\
  \hat{0} & \hat{G}^-_k \end{pmatrix}
  \label{eq:G4by4}
\end{equation}
with $G^{+(-)}_k$ being the retarded (advanced)
Green's function
and
\begin{equation}
  \hat{G}^K_k={\rm tanh}\left(\frac{\epsilon}{2T}\right)
  \left(\hat{G}^+_k-\hat{G}^-_k\right),
  \label{eq:Gkeldysh}
\end{equation}
where $T$ is the temperature and 
$k=({\mib k},\epsilon)$.

We consider that $G$ includes
the mean-field superconducting gap ($\Delta$)
given by the electron--phonon interaction
and 
the effect of the isotropic impurity scattering with the Born approximation,
and $V$ describes other interaction effects.
The one-particle Green's function
is written as~\cite{memo1,memo2}
\begin{equation}
\hat{G}^{\pm}_k=
\frac{\eta_{\epsilon}^{\pm}\epsilon\hat{\tau}_0
  +\xi_{\mib k}\hat{\tau}_3+\eta_{\epsilon}^{\pm}\Delta\hat{\tau}_1}
{(\eta_{\epsilon}^{\pm}\epsilon)^2
  -\xi_{\mib k}^2-(\eta_{\epsilon}^{\pm}\Delta)^2}.
\label{eq:Gretadv}
\end{equation}
Here,
$\hat{\tau}_0=\left(\begin{smallmatrix}1&0\\0&1\end{smallmatrix}\right)$,
$\hat{\tau}_3=\left(\begin{smallmatrix}1&0\\0&-1\end{smallmatrix}\right)$,
$\hat{\tau}_1=\left(\begin{smallmatrix}0&1\\1&0\end{smallmatrix}\right)$,
and
$\Delta$ is determined by the gap equation
\begin{equation}
  \Delta\hat{\tau}_1=
  \frac{g_{ph}^2}
       {\omega_EN^3}\sum_{\mib k}\int\frac{d\epsilon}{2\pi i}
  \hat{\tau}_3\hat{G}^K_k\hat{\tau}_3.
\end{equation}
$\eta_{\epsilon}^{\pm}$ is given by the following
equation including the impurity scattering:
\begin{equation}
  \eta_{\epsilon}^{\pm}\epsilon\hat{\tau}_0
  -\eta_{\epsilon}^{\pm}\Delta\hat{\tau}_1
  =
  \epsilon\hat{\tau}_0-\Delta\hat{\tau}_1
  -\frac{n_i u^2}{N^3}\sum_{\mib k}\hat{\tau}_3\hat{G}^{\pm}_k
  \hat{\tau}_3,
\end{equation}
and is written as
\begin{equation}
  \eta_{\epsilon}^{\pm}=1+\frac{\alpha}{\zeta_{\epsilon}^{\pm}}.
\end{equation}
$\alpha:=n_i u^2m k_F/2\pi=1/2\tau$ with $n_i$ and $m$ being
the concentration of impurities and
the mass of quasiparticles, respectively,
with
\begin{equation}
  \zeta^{\pm}_{\epsilon}=
  -i{\rm sgn}(\epsilon)\sqrt{\epsilon^2-\Delta^2}\theta(|\epsilon|-\Delta)
  +\sqrt{\Delta^2-\epsilon^2}\theta(\Delta-|\epsilon|).
\end{equation}
$\theta(\cdot)$ is a step function.
$V=V_A+V_i+V_C+V_p$ describes the coupling
to the external field ($V_A$) and 
vertex corrections
in the conductivity. 
$V_C$, $V_p$, and $V_i$
represent 
the Coulomb interaction between electrons,
the electron--phonon interaction,
and the scattering by impurities, respectively.
These are written as 
\begin{equation}
  V_A=
  \frac{e
    {\mib v}_{(\mib k+\mib k')/2}}{\sqrt{2}\sqrt{2\pi N^3}}
  \cdot
   \begin{pmatrix}
      \hat{\tau}_0\mib{A}^{cl}_{k-k'}&\hat{\tau}_0\mib{A}^{qu}_{k-k'}\\
      \hat{\tau}_0\mib{A}^{qu}_{k-k'}&\hat{\tau}_0\mib{A}^{cl}_{k-k'}
   \end{pmatrix},
   \label{eq:VA4by4}
\end{equation}
\begin{equation}
  V_C=\frac{-i e}{\sqrt{2}\sqrt{2\pi N^3}}
  \begin{pmatrix}\hat{\tau}_3\varphi^{cl}_{k-k'}
    &\hat{\tau}_3\varphi^{qu}_{k-k'}
    \\\hat{\tau}_3\varphi^{qu}_{k-k'}
    &\hat{\tau}_3\varphi^{cl}_{k-k'}  \end{pmatrix},
     \label{eq:VC4by4}
\end{equation}
\begin{equation}
  V_{p}=\frac{-g_{ph}}
{\sqrt{2}\sqrt{2\pi N^3}}
 \begin{pmatrix}\hat{\tau}_3\phi^{cl}_{k-k'}
   &\hat{\tau}_3\phi^{qu}_{k-k'}
   \\\hat{\tau}_3\phi^{qu}_{k-k'}
   &\hat{\tau}_3\phi^{cl}_{k-k'}\end{pmatrix},
    \label{eq:Vp4by4}
\end{equation}
and 
\begin{equation}
  V_i=\frac{-u_{\mib k-\mib k'}}{\sqrt{2\pi N^3}}
  \begin{pmatrix} 
    \hat{\tau}_3 & \hat{0}
    \\ \hat{0}&\hat{\tau}_3  \end{pmatrix}
  \sqrt{2\pi}\delta(\epsilon-\epsilon').
     \label{eq:Vi4by4}
\end{equation}
$\delta(\cdot)$ is a delta function.

\section{Expressions for Correction Terms}

In this section, we show expressions for
the linear absorption
including vertex corrections.
A detailed derivation of these expressions is given in
Appendix.
The following subsection shows a method of
calculating four-point interaction vertices.

\subsection{Vertex corrections}

\begin{figure}
    \includegraphics[width=11.5cm]{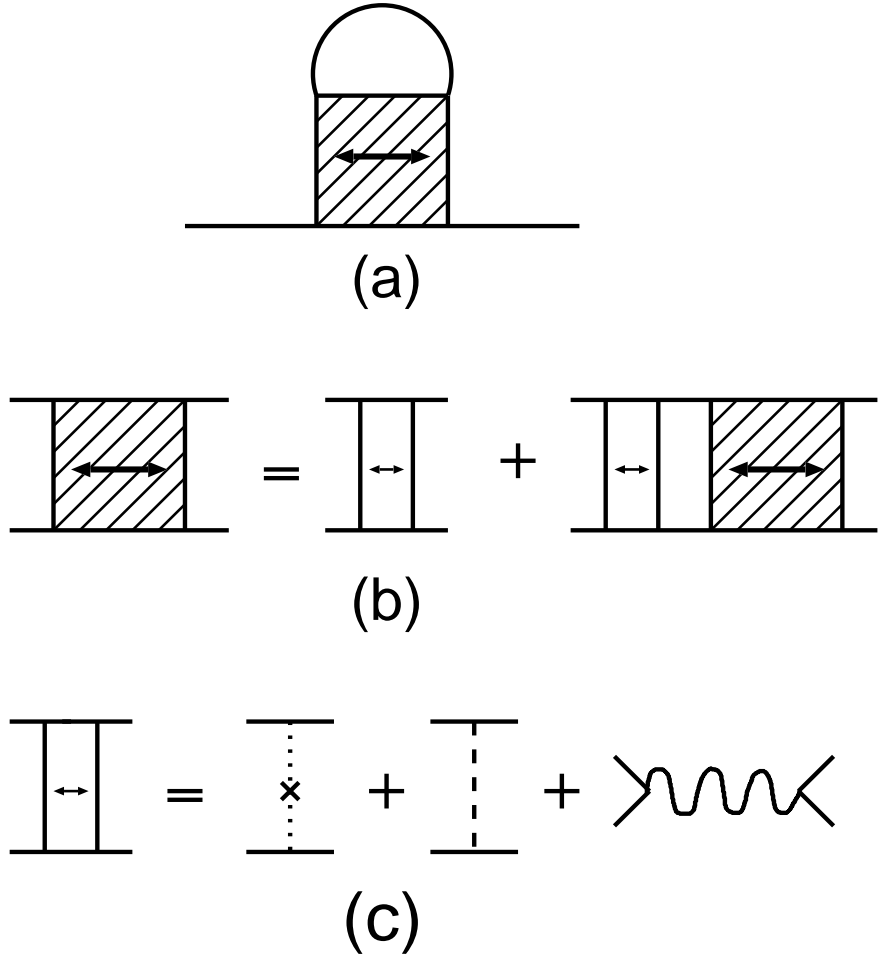}
    \caption{\label{fig:1}
      (a) Diagram of the weak localization correction to the one-particle
      spectrum. The effect of interactions is included in
      the shaded square (the interaction vertex).
      The solid lines indicate the one-particle Green function
      [Eqs. (\ref{eq:G4by4})-(\ref{eq:Gretadv})].
      (b) Diagram of the interaction vertex, which includes
      the screened Coulomb interaction, superconducting fluctuation,
      and diffuson. The arrow in the square specifies the
      direction of the interaction vertex. 
      (c) The bare interaction vertex consists of
      the bare Coulomb interaction (wavy line),
      electron--phonon interaction (dashed line),
      and impurity scattering (dotted line with a cross).
    }
\end{figure}

The coefficient of $A^{cl}_q A^{qu}_{-q}$ in 
$\langle exp[\sum_n (-1)^n{\rm tr}(G V)^n/n]\rangle_{\varphi,p,i}$
gives the linear response including vertex corrections.
We take this vertex correction to be consistent
with the weak localization correction to the one-particle
spectrum.~\cite{jujo2019}
The diagram of the latter correction is shown in Fig.~\ref{fig:1}.
The weak localization effect mainly arises from
the screened Coulomb interaction corrected by the diffuson. 
The superconducting fluctuation is taken into account because
of the coupling between the electron density and the
phase of the superconducting order parameter.

\begin{figure}
  \includegraphics[width=9.5cm]{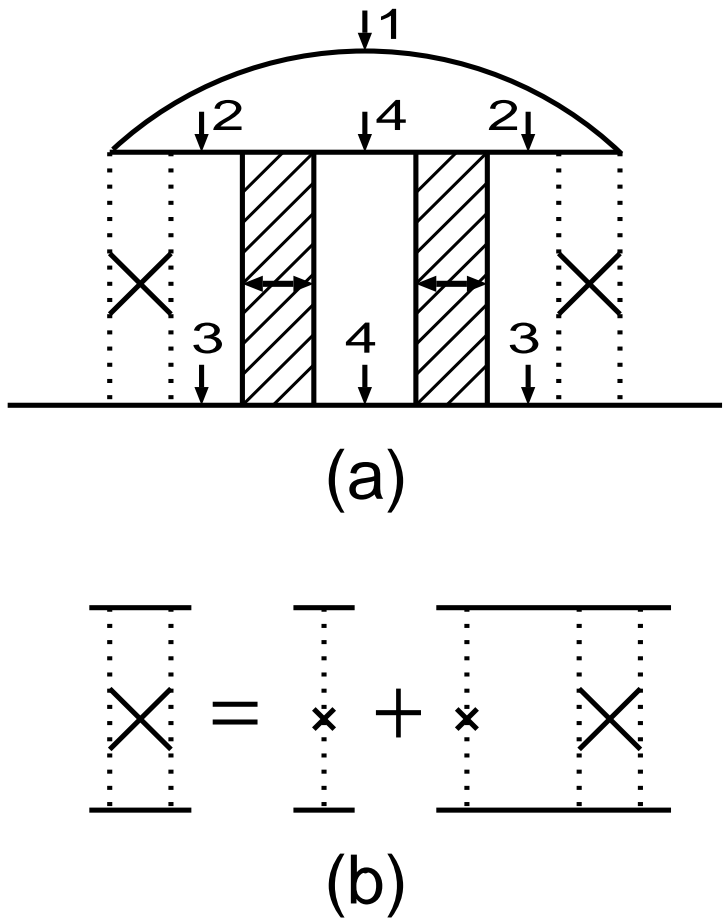}
  \caption{\label{fig:2}
    (a) The self-energy correction is rewritten
    to obtain the vertex correction to the response function.
    The solid lines indicate the one-particle Green function
    [Eqs. (\ref{eq:G4by4})--(\ref{eq:Gretadv})].
    The meaning of the shaded square is the same as that
    in Fig. 1.
    The dotted square including a cross means the
    diffuson propagator.
    The arrow numbers indicate the corresponding vertex corrections.
    (b) Diffuson propagator. This is given by the ladder
    of the impurity scattering, which is represented by
    a single dotted line with a cross.
  }
\end{figure}

The vertex corrections to the response function are given by the
functional derivative of the self-energy by the one-particle
Green function.~\cite{baym}
These four-point interaction vertices are obtained by cutting the
lines specified by the arrows in the self-energy in Fig.~\ref{fig:2}(a),
which is equivalent to Fig. 1(a).
The arrow numbers indicate the corresponding vertex corrections.
When we cut the solid lines with arrow numbers ``1'', ``2'', ``3'', and ``4'',
we obtain the four-point interaction vertices corresponding to
the Maki--Thompson (MT) term,
the MT term corrected by the diffuson,
the density of states (DOS) term,
and the Aslamazov--Larkin (AL) term, respectively.
These are written as ``$MT0$'', ``$MT$'', ``$DOS$'', and ``$AL$''
below, respectively.

The four-point interaction vertex obtained in this way
is called an irreducible four-point interaction vertex.
The reducible four-point interaction vertex is obtained
by the ladder-type summation of the irreducible vertex.
As the number of rungs of the ladder increases,
the exponent $n$ of $1/(k_Fl)^{2n}$
(the coefficient of the correction term) 
increases.
Thus, we take account of only 
the lowest order of $1/(k_Fl)^2$.~\cite{memo3}
  This perturbation method is consistent with an approximation
  that the one-particle Green's function
  [Eqs. (\ref{eq:G4by4})--(\ref{eq:Gretadv})] does not include
  the weak localization effect,~\cite{memo1}
  and it is necessary to
  calculate the self-energy term
  in Sect. 3.2.2 [Fig. 4(a)] below
  because of this approximation.

\subsection{Expressions of correction terms for linear absorption}

\subsubsection{Maki--Thompson terms}

\begin{figure}
  \includegraphics[width=11.5cm]{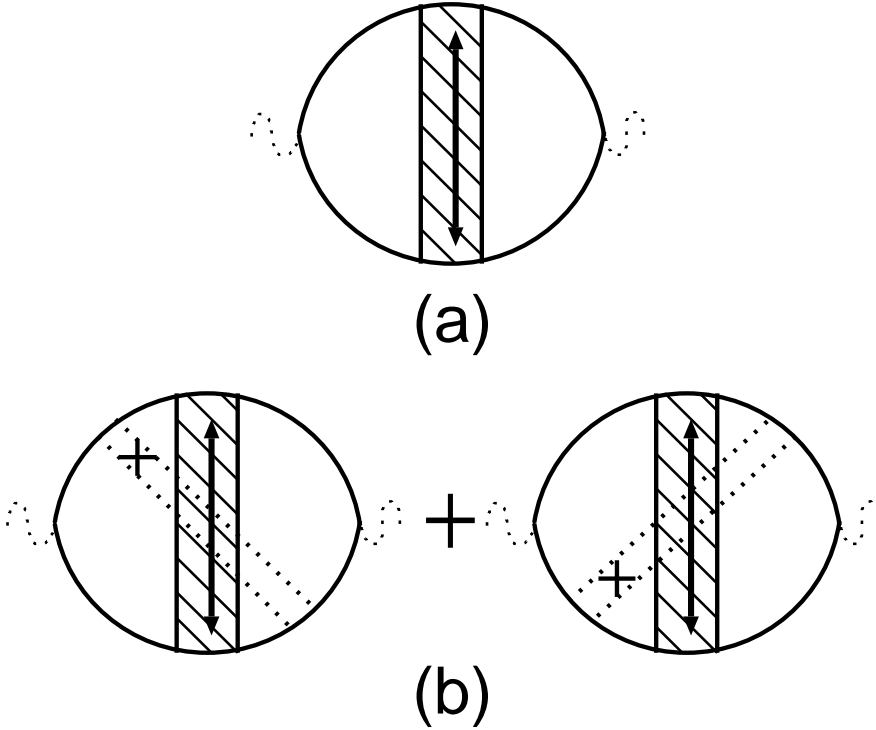}
  \caption{\label{fig:3}
    (a) Diagram of the conductivity corrected
    by the four-point interaction vertex, which is obtained by cutting the line
    ``1'' in Fig. 2(a) ($\sigma^{MT0}_{\omega}$).
    (b) Diagram of the conductivity
    derived in the same way as in (a) by
    cutting the line ``2'' in Fig. 2(a) ($\sigma^{MT}_{\omega}$).
    The dotted wavy line indicates the current vertex. 
  }
\end{figure}

The diagrams of the MT terms are shown in
Fig.~\ref{fig:3}.
As derived in Appendix A.1,
the expression for one of the MT terms [Fig. 3(a)] is
written as 
\begin{equation}
  \frac{{\rm Re}\sigma^{MT0}_{\omega}}{\sigma_0}
  =
  \frac{-3\sqrt{3\tau}}{\omega(4\pi k_Fl)^2}
  \int d x\sqrt{x}\int d\epsilon \int d\omega'
       {\rm Im}Q^{MT0}_{\epsilon,\omega',x}(\omega)
       \label{eq:condMT0}
\end{equation}
($x=D q'^2$ with $D=v_F^2\tau/3$ being the diffusion constant
and $\sigma_0=e^2n_e\tau/m$ with $n_e=k_F^3/3\pi^2$).
Here,
\begin{equation}
  \begin{split}
    Q^{MT0}_{\epsilon,\omega',x}(\omega)=& \sum_{i=0,1,2,3,4}
    2\Gamma_i(q')\{
  C^t_{\omega'}\sum_{s=\pm}s(T^h_{\epsilon_4}{\cal N}_i^{+++s}
  +T^h_{\epsilon_3}{\cal N}_i^{++s-}+T^h_{\epsilon_2}{\cal N}_i^{+s--}
  +T^h_{\epsilon_1}{\cal N}_i^{s---})\\
&  +\sum_{s,s'=\pm}s s'(T^h_{\epsilon_3}T^h_{\epsilon_4}{\cal N}_i^{++s s'}
  +T^h_{\epsilon_2}T^h_{\epsilon_4}{\cal N}_i^{+s-s'}
  )
  +{\cal N}_i^{+++-}+{\cal N}_i^{---+}\}
  \end{split}
  \label{eq:MT0cdot}
\end{equation}
with $C^t_{\omega}={\rm coth}(\omega/2T)$,
$T^h_{\epsilon}={\rm tanh}(\epsilon/2T)$,
$\epsilon_1=\epsilon+(\omega+\omega')/2$,
$\epsilon_2=\epsilon+(\omega-\omega')/2$,
$\epsilon_3=\epsilon-(\omega+\omega')/2$, and 
$\epsilon_4=\epsilon-(\omega-\omega')/2$.
\begin{equation}
  {\cal N}_i^{s_1s_2s_3s_4}=
  \frac{
    {\rm Tr}[
            (\hat{\tau}_0+s\hat{\tau}_j\hat{g}^{s_1}_{\epsilon_1}
    \hat{\tau}_j\hat{g}^{s_2}_{\epsilon_2})
    (\hat{\tau}_0+s\hat{g}^{s_3}_{\epsilon_3}
    \hat{\tau}_j\hat{g}^{s_4}_{\epsilon_4}    \hat{\tau}_j)
    ]}
              {2(x+\zeta^{s_1}_{\epsilon_1}+\zeta^{s_2}_{\epsilon_2})
    (x+\zeta^{s_3}_{\epsilon_3}+\zeta^{s_4}_{\epsilon_4})  }
\end{equation}
for $i=0,1,2,3$
$\left[
s=\left\{
 \begin{smallmatrix} 1 & (i=0,3) \\ -1 & (i=1,2) \end{smallmatrix}
 \right.
      \text{ and }
j=\left\{
 \begin{smallmatrix} 3 & (i=2,3) \\ 0 & (i=0,1) \end{smallmatrix}
 \right.      
\right]$      
and
\begin{equation}
  {\cal N}_{4}^{s_1s_2s_3s_4}=
  \frac{-
    {\rm Tr}[
\hat{\tau}_1
    \hat{g}^{s_1}_{\epsilon_1}
    \hat{\tau}_j\hat{g}^{s_2}_{\epsilon_2}
    \hat{g}^{s_3}_{\epsilon_3}
    \hat{\tau}_j\hat{g}^{s_4}_{\epsilon_4}]}
              {(x+\zeta^{s_1}_{\epsilon_1}+\zeta^{s_2}_{\epsilon_2})
    (x+\zeta^{s_3}_{\epsilon_3}+\zeta^{s_4}_{\epsilon_4})  }
\end{equation}
with
$\hat{g}_{\epsilon}^{\pm}:=g_{\epsilon}^{\pm}\hat{\tau}_0+
f_{\epsilon}^{\pm}\hat{\tau}_1$ 
($g_{\epsilon}^{\pm}=-\epsilon/\zeta^{\pm}_{\epsilon}$
and
$f_{\epsilon}^{\pm}=-\Delta/\zeta^{\pm}_{\epsilon}$).
$\Gamma_{i=0,1,2,3,4}$ represent the screened Coulomb interaction
and the superconducting fluctuation~\cite{memo4}
and are written as 
\begin{equation}
  \begin{pmatrix}
  \Gamma_3(q)\\
  \Gamma_2(q)\\
  \Gamma_{4}(q)
  \end{pmatrix}
  =
  \frac{1}{
(1/p+\chi_2)[1-(p+c_q)\chi_3]+4(p+c_q)(\chi')^2
  }
  \begin{pmatrix}
    (p+c_q)(1/p+\chi_2)/2 \\
    -[1-(p+c_q)\chi_3]/2 \\
    -(p+c_q)\chi'
  \end{pmatrix},
  \label{eq:Gam32d}
\end{equation}
\begin{equation}
  \Gamma_0(q)=\frac{p/2}{1-p\chi_0},
  \label{eq:Gam0}
\end{equation}
and
\begin{equation}
  \Gamma_1(q)=\frac{-1/2}{1/p+\chi_1}
  \label{eq:Gam1}
\end{equation}
with $p:=(\pi\rho_0/2)(g_{ph}^2/\omega_E)$ and 
$c_q:=(\pi\rho_0/2)v^C_{\mib q}$ ($\rho_0:=m k_F/\pi^2$). 
\begin{equation}
  \chi_i=\sum_{s=\pm}s\int\frac{d\epsilon}{2\pi i}
 {\rm Tr} \left[
  \frac{T^h_{\epsilon}X_{\epsilon+\omega,\epsilon}^{+s}
    \hat{\tau}_i(h_i\hat{\tau}_i+
         \hat{g}^{+}_{\epsilon+\omega}\hat{\tau}_i\hat{g}^{s}_{\epsilon})}
       {2\alpha(1-2X_{\epsilon+\omega,\epsilon}^{+s})}
+  
\frac{T^h_{\epsilon+\omega}X_{\epsilon+\omega,\epsilon}^{s-}
  \hat{\tau}_i(h_i\hat{\tau}_i+
         \hat{g}^{s}_{\epsilon+\omega}\hat{\tau}_i\hat{g}^{-}_{\epsilon})}
       {2\alpha(1-2X_{\epsilon+\omega,\epsilon}^{s-})}
       \right]
       -\frac{2}{\pi}h_i''
\end{equation}
for $i=0,1,2,3$
$\left[  h_i=\left\{\begin{smallmatrix} 1 & (i=0,3) \\ -1 & (i=1,2)
  \end{smallmatrix}\right.
  \text{ and }
  h''_i=\left\{\begin{smallmatrix} 1 & (i=0,3) \\ 0 & (i=1,2)
  \end{smallmatrix}\right.
\right]$,  
and
\begin{equation}
  \chi{'}=\sum_{s=\pm}s\int\frac{d\epsilon}{2\pi i}
       {\rm Tr} \left[
  \frac{T^h_{\epsilon}X_{\epsilon+\omega,\epsilon}^{+s}
    (-i\hat{\tau}_2)
         \hat{g}^{+}_{\epsilon+\omega}\hat{\tau}_3\hat{g}^{s}_{\epsilon}}
       {4\alpha(1-2X_{\epsilon+\omega,\epsilon}^{+s})}
+  
\frac{T^h_{\epsilon+\omega}X_{\epsilon+\omega,\epsilon}^{s-}
  (-i\hat{\tau}_2)
         \hat{g}^{s}_{\epsilon+\omega}\hat{\tau}_3\hat{g}^{-}_{\epsilon}}
       {4\alpha(1-2X_{\epsilon+\omega,\epsilon}^{s-})}
       \right]
\end{equation}
[$\hat{\tau}_2=\left(\begin{smallmatrix}0&-i\\i&0\end{smallmatrix}\right)$
].

The expression of the real part of the conductivity corresponding to Fig. 3(b)
(the MT term with an additional diffuson)
is given by the following equation, the derivation of which is given
in Appendix A.1:
\begin{equation}
  \frac{{\rm Re}\sigma^{MT}_{\omega}}{\sigma_0}
  =  \frac{\sqrt{3\tau}}{\omega(4\pi k_Fl)^2}
  \int d x x^{3/2}\int d\epsilon \int d\omega'
       {\rm Im}Q^{MT}_{\epsilon,\omega',x}(\omega).
       \label{eq:condMT}
\end{equation}
$Q^{MT}_{\epsilon,\omega',x}(\omega)$ is given by
Eq. (\ref{eq:MT0cdot}) with ${\cal N}$ replaced
by the following ${\cal M}$:
\begin{equation}
  \begin{split}
  {\cal M}_i^{s_1s_2s_3s_4}=&
    \frac{1}
              {(x+\zeta^{s_1}_{\epsilon_1}+\zeta^{s_2}_{\epsilon_2})
                (x+\zeta^{s_3}_{\epsilon_3}+\zeta^{s_4}_{\epsilon_4})  }
         \left(\frac{1}{x+\zeta^{s_1}_{\epsilon_1}+\zeta^{s_3}_{\epsilon_3}}
 +\frac{1}{x+\zeta^{s_2}_{\epsilon_2}+\zeta^{s_4}_{\epsilon_4}}
 \right)\\
 &\times
 \frac{1}{2}{\rm Tr}[
            (\hat{\tau}_0+s\hat{\tau}_j\hat{g}^{s_1}_{\epsilon_1}
    \hat{\tau}_j\hat{g}^{s_2}_{\epsilon_2})
(\hat{\tau}_0    +s\hat{g}^{s_3}_{\epsilon_3}
    \hat{\tau}_j\hat{g}^{s_4}_{\epsilon_4}    \hat{\tau}_j)
    +
            (\hat{\tau}_{j'}\hat{g}^{s_1}_{\epsilon_1}
    \hat{\tau}_{j'}-s\hat{g}^{s_2}_{\epsilon_2})
(s\hat{g}^{s_3}_{\epsilon_3}-
    \hat{\tau}_{j'}\hat{g}^{s_4}_{\epsilon_4}    \hat{\tau}_{j'})    
        ]
  \end{split}
  \label{eq:calM}
  \end{equation}
for $i=0,1,2,3$
$\left[
s=\left\{
 \begin{smallmatrix} 1 & (i=0,3) \\ -1 & (i=1,2) \end{smallmatrix}
 \right.,
 \text{ }
j=\left\{
 \begin{smallmatrix} 3 & (i=2,3) \\ 0 & (i=0,1) \end{smallmatrix}
 \right.      
\text{ and }
j'=\left\{
 \begin{smallmatrix} 0 & (i=2,3) \\ 3 & (i=0,1) \end{smallmatrix}
 \right.      
\right]$      
and
\begin{equation}
  {\cal M}_{4}^{s_1s_2s_3s_4}=
    \frac{{\rm Tr}[
        \hat{\tau}_1
                    (\hat{g}^{s_1}_{\epsilon_1}\hat{g}^{s_4}_{\epsilon_4}
-\hat{g}^{s_2}_{\epsilon_2}\hat{g}^{s_3}_{\epsilon_3}
-    \hat{g}^{s_1}_{\epsilon_1}
    \hat{\tau}_j\hat{g}^{s_2}_{\epsilon_2}
    \hat{g}^{s_3}_{\epsilon_3}
    \hat{\tau}_j\hat{g}^{s_4}_{\epsilon_4})]}
              {(x+\zeta^{s_1}_{\epsilon_1}+\zeta^{s_2}_{\epsilon_2})
                (x+\zeta^{s_3}_{\epsilon_3}+\zeta^{s_4}_{\epsilon_4})  }
         \left(\frac{1}{x+\zeta^{s_1}_{\epsilon_1}+\zeta^{s_3}_{\epsilon_3}}
         +\frac{1}{x+\zeta^{s_2}_{\epsilon_2}+\zeta^{s_4}_{\epsilon_4}}
         \right).
\end{equation}

\subsubsection{Density of states terms}

\begin{figure}
  \includegraphics[width=11.5cm]{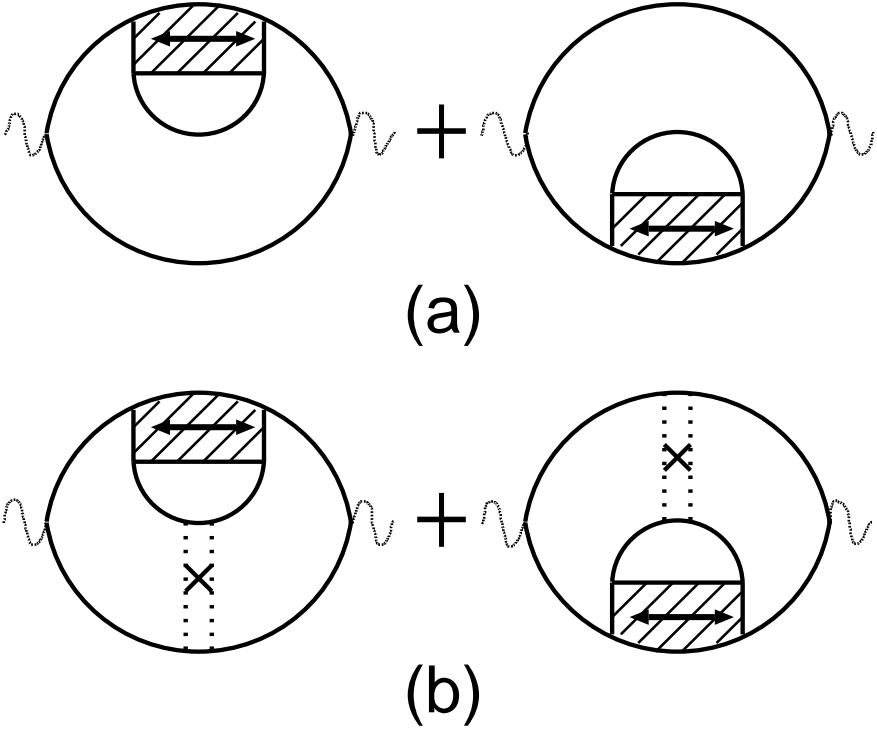}
  \caption{\label{fig:4}
    (a) Diagram of the conductivity
    corrected by the self-energy term ($\sigma^{DOS0}_{\omega}$).
    (b) Diagram of the conductivity corrected
    by the four-point interaction vertex, which is obtained by cutting the line
    ``3'' in Fig. 2(a) ($\sigma^{DOS}_{\omega}$).
  }
\end{figure}

The diagrams of the DOS terms are shown in
Fig.~\ref{fig:4}.
As derived in Appendix A.2,
the result of the self-energy term [Fig. 4(a)] is written as 
\begin{equation}
  \frac{{\rm Re}\sigma^{DOS0}_{\omega}}{\sigma_0}
  =
  \frac{-3\sqrt{3\tau}}{\omega(8\pi k_Fl)^2}
  \int d x\sqrt{x}\int d\epsilon \int d\omega'
       {\rm Im}Q^{DOS0}_{\epsilon,\omega',x}(\omega)
       \label{eq:condDOS0}
\end{equation}
with
\begin{equation}
  \begin{split}
    &  Q^{DOS0}_{\epsilon,\omega',x}(\omega)= \sum_{i=0,1,2,3,4}2
\{
    C^t_{\omega'}[\Gamma_i(q')-\Gamma_i^*(q')]
    \sum_{s=\pm}s(T^h_{\epsilon_4}{\cal S}_i^{+++s}
    +T^h_{\epsilon_1}{\cal S}_i^{s s s-})\\    
  &  +\Gamma_i(q')(
    \sum_{s,s'=\pm}s s' T^h_{\epsilon_2}T^h_{\epsilon_4}{\cal S}_i^{+s+s'}
    +  \sum_{s=\pm}s T^h_{\epsilon_2}T^h_{\epsilon_1}{\cal S}_i^{+s+-}
)
  -\Gamma_i^*(q')
      \sum_{s=\pm}s T^h_{\epsilon_1}T^h_{\epsilon_2}{\cal S}_i^{-s--}
\},
  \end{split}
  \label{eq:DOS0cdot}
\end{equation}
where $\Gamma^*$ means the complex conjugate of $\Gamma$.
Here,
\begin{equation}
  {\cal S}_i^{s_1s_2s_3s_4}=
  \frac{
    {\rm Tr}[
      s\hat{\tau}_0+\hat{g}^{s_1}_{\epsilon_1}\hat{\tau}_j
      \hat{g}^{s_2}_{\epsilon_2}\hat{\tau}_j
      -s\hat{g}^{s_1}_{\epsilon_1}\hat{\tau}_3
      \hat{g}^{s_3}_{\epsilon_1}\hat{\tau}_3
      +\hat{g}^{s_2}_{\epsilon_2}\hat{\tau}_j
      \hat{g}^{s_3}_{\epsilon_1}\hat{\tau}_j
      +3(s\hat{g}^{s_1}_{\epsilon_1}
      -\hat{\tau}_{j'}\hat{g}^{s_2}_{\epsilon_2}\hat{\tau}_{j'}
      +s\hat{g}^{s_3}_{\epsilon_1}+
      \hat{g}^{s_1}_{\epsilon_1}\hat{\tau}_j\hat{g}^{s_2}_{\epsilon_2}
      \hat{\tau}_j
      \hat{g}^{s_3}_{\epsilon_1})\hat{g}^{s_4}_{\epsilon_4}
]  }{2(x+\zeta^{s_1}_{\epsilon_1}+\zeta^{s_2}_{\epsilon_2})
(x+\zeta^{s_3}_{\epsilon_1}+\zeta^{s_2}_{\epsilon_2})}
\end{equation}
for $i=0,1,2,3$
[the values of $s$, $j$, and $j'$ are
the same as those in ${\cal M}$ below Eq. (\ref{eq:calM})]
and
\begin{equation}
  {\cal S}_{4}^{s_1s_2s_3s_4}=
  \frac{-{\rm Tr}[
      \hat{\tau}_1(
      \hat{g}^{s_1}_{\epsilon_1}\hat{\tau}_3
      \hat{g}^{s_2}_{\epsilon_2}\hat{\tau}_3
      -\hat{g}^{s_2}_{\epsilon_2}\hat{\tau}_3
      \hat{g}^{s_3}_{\epsilon_1}\hat{\tau}_3
      +3\hat{g}^{s_2}_{\epsilon_2}\hat{g}^{s_4}_{\epsilon_4}
      +3
      \hat{g}^{s_1}_{\epsilon_1}\hat{\tau}_3\hat{g}^{s_2}_{\epsilon_2}
      \hat{\tau}_3
      \hat{g}^{s_3}_{\epsilon_1}\hat{g}^{s_4}_{\epsilon_4}
)]  }{(x+\zeta^{s_1}_{\epsilon_1}+\zeta^{s_2}_{\epsilon_2})
(x+\zeta^{s_3}_{\epsilon_1}+\zeta^{s_2}_{\epsilon_2})}.
\end{equation}

The expression of the DOS term shown in Fig. 4(b)
is given by the following equation:
\begin{equation}
  \frac{{\rm Re}\sigma^{DOS}_{\omega}}{\sigma_0}
  =
  \frac{-\sqrt{3\tau}}{\omega(4\pi k_Fl)^2}
  \int d x x^{3/2}\int d\epsilon \int d\omega'
       {\rm Im}Q^{DOS}_{\epsilon,\omega',x}(\omega).
       \label{eq:condDOS}
\end{equation}
$Q^{DOS}_{\epsilon,\omega',x}(\omega)$ is given by Eq. (\ref{eq:DOS0cdot})
with ${\cal S}$ replaced
by the following ${\cal D}$:
\begin{equation}
  {\cal D}_i^{s_1s_2s_3s_4}=
  \frac{
    {\rm Tr}[
      s\hat{\tau}_0+\hat{g}^{s_1}_{\epsilon_1}\hat{\tau}_j
      \hat{g}^{s_2}_{\epsilon_2}\hat{\tau}_j
      -s\hat{g}^{s_1}_{\epsilon_1}\hat{\tau}_3
      \hat{g}^{s_3}_{\epsilon_1}\hat{\tau}_3
      +\hat{g}^{s_2}_{\epsilon_2}\hat{\tau}_j
      \hat{g}^{s_3}_{\epsilon_1}\hat{\tau}_j
      -(s\hat{g}^{s_1}_{\epsilon_1}
      -\hat{\tau}_{j'}\hat{g}^{s_2}_{\epsilon_2}\hat{\tau}_{j'}
      +s\hat{g}^{s_3}_{\epsilon_1}+
      \hat{g}^{s_1}_{\epsilon_1}\hat{\tau}_j\hat{g}^{s_2}_{\epsilon_2}
      \hat{\tau}_j
      \hat{g}^{s_3}_{\epsilon_1})\hat{g}^{s_4}_{\epsilon_4}
]  }{2(x+\zeta^{s_1}_{\epsilon_1}+\zeta^{s_2}_{\epsilon_2})
    (x+\zeta^{s_3}_{\epsilon_1}+\zeta^{s_2}_{\epsilon_2})
    (x+\zeta^{s_4}_{\epsilon_4}+\zeta^{s_2}_{\epsilon_2})}
\end{equation}
for $i=0,1,2,3$
(the values of $s$, $j$, and $j'$ are the same as those
      in the cases of ${\cal M}$ and ${\cal S}$ above)
and
\begin{equation}
  {\cal D}_{4}^{s_1s_2s_3s_4}=
  \frac{-{\rm Tr}[
      \hat{\tau}_1(
      \hat{g}^{s_1}_{\epsilon_1}\hat{\tau}_3
      \hat{g}^{s_2}_{\epsilon_2}\hat{\tau}_3
      -\hat{g}^{s_2}_{\epsilon_2}\hat{\tau}_3
      \hat{g}^{s_3}_{\epsilon_1}\hat{\tau}_3
      -\hat{g}^{s_2}_{\epsilon_2}\hat{g}^{s_4}_{\epsilon_4}
      -
      \hat{g}^{s_1}_{\epsilon_1}\hat{\tau}_3\hat{g}^{s_2}_{\epsilon_2}
      \hat{\tau}_3
      \hat{g}^{s_3}_{\epsilon_1}\hat{g}^{s_4}_{\epsilon_4}
)]  }{(x+\zeta^{s_1}_{\epsilon_1}+\zeta^{s_2}_{\epsilon_2})
    (x+\zeta^{s_3}_{\epsilon_1}+\zeta^{s_2}_{\epsilon_2})
    (x+\zeta^{s_4}_{\epsilon_4}+\zeta^{s_2}_{\epsilon_2})}.
\end{equation}

\subsubsection{Aslamazov--Larkin term}

\begin{figure}
  \includegraphics[width=11.5cm]{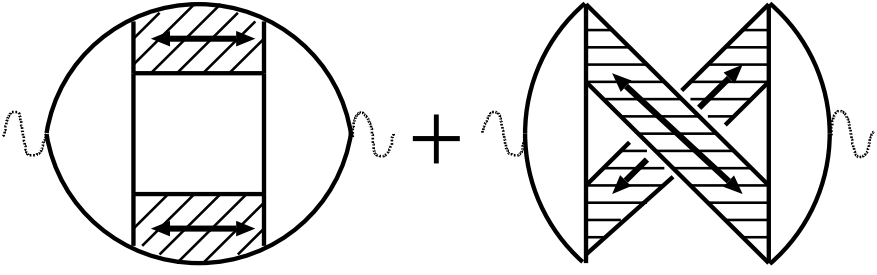}
  \caption{\label{fig:5}
    Diagram of the conductivity corrected
    by the four-point interaction vertex obtained by cutting ``4'' lines
    in Fig. 2(a). $\sigma^{AL}_{\omega}
    =\sigma^{AL1}_{\omega}+\sigma^{AL2}_{\omega}$
    with the left diagram ($\sigma^{AL1}_{\omega}$)
    and the right diagram ($\sigma^{AL2}_{\omega}$).
}
\end{figure}

The diagram of the AL term is shown in 
Fig.~\ref{fig:5}.
The result of the AL term ($\sigma^{AL}_{\omega}
=\sigma^{AL1}_{\omega}+\sigma^{AL2}_{\omega}$)
is given by the following equation, the derivation of which is
given in Appendix A.3:
\begin{equation}
  \frac{  {\rm Re}\sigma^{AL}_{\omega}}{\sigma_0}
  =
  \frac{\sqrt{3\tau}}{2\pi^3\omega(k_Fl)^2}
  \int d\omega'\int d x x^{3/2}
       {\rm Re}Q^{AL}_{\omega',x}(\omega)
       \label{eq:condAL}
\end{equation}
with
\begin{equation}
  \begin{split}
Q^{AL}_{\omega',x}(\omega)=&\sum_{i=3,2}\{
\Gamma_1(q-q')[  C^t_{\omega'}
    \Gamma_i(q')
  ({\cal A}^{(1)}_{i,1})^2
  +(C^t_{\omega}-C^t_{\omega'})\Gamma^*_i(q')
    ({\cal A}^{(2)}_{i,1})^2
  -    C^t_{\omega}
  \Gamma^*_i(q'){\cal A}^{(2)}_{i,1}{\cal A}^{(3)}_{i,1}]
  \\
  &+
  \Gamma_i(q-q')[  C^t_{\omega'}
    \Gamma_1(q')
  ({\cal A}^{(1)}_{i,2})^2
  +(C^t_{\omega}-C^t_{\omega'})\Gamma^*_1(q')
    ({\cal A}^{(2)}_{i,2})^2
  -    C^t_{\omega}
  \Gamma^*_1(q'){\cal A}^{(2)}_{i,2}{\cal A}^{(3)}_{i,2}
  ]\}
  \\
  &+
  \Gamma_1(q-q')[  2C^t_{\omega'}
    \Gamma_{4}(q')
  {\cal A}^{(1)}_{2,1}{\cal A}^{(1)}_{3,1}
  +2(C^t_{\omega}-C^t_{\omega'})\Gamma^*_{4}(q')
    {\cal A}^{(2)}_{2,1}{\cal A}^{(2)}_{3,1}
  -    C^t_{\omega}
  \Gamma^*_{4}(q')
  ({\cal A}^{(2)}_{2,1}{\cal A}^{(3)}_{3,1}
  +{\cal A}^{(2)}_{3,1}{\cal A}^{(3)}_{2,1})
  ]
  \\
  &-
  \Gamma_{4}(q-q')[  2C^t_{\omega'}
    \Gamma_1(q')
  {\cal A}^{(1)}_{2,2}{\cal A}^{(1)}_{3,2}
  +2(C^t_{\omega}-C^t_{\omega'})\Gamma^*_1(q')
    {\cal A}^{(2)}_{2,2}{\cal A}^{(2)}_{3,2}
  -    C^t_{\omega}
  \Gamma^*_1(q')
  ({\cal A}^{(2)}_{2,2}{\cal A}^{(3)}_{3,2}
  +{\cal A}^{(2)}_{3,2}{\cal A}^{(3)}_{2,2})
  ].
  \end{split}
  \label{eq:QAL}
\end{equation}
Here,
\begin{equation}
{\cal A}^{(1)}_{i,j}=\sum_{s=\pm}
  s\int d\epsilon
  (T^h_{\epsilon}{\cal L}_{i,j}^{+s+}
  +T^h_{\epsilon+\omega'}{\cal L}_{i,j}^{+-s}
  +T^h_{\epsilon+\omega}{\cal L}_{i,j}^{s--}),
\end{equation}
\begin{equation}
{\cal A}^{(2)}_{i,j}=\sum_{s=\pm}
  s\int d\epsilon
  (T^h_{\epsilon+\omega'}{\cal L}_{i,j}^{++s}
  +T^h_{\epsilon}{\cal L}_{i,j}^{+s-}
  +T^h_{\epsilon+\omega}{\cal L}_{i,j}^{s--}),
\end{equation}
and
\begin{equation}
  {\cal A}^{(3)}_{i,j}=
  \sum_{s=\pm}
  s\int d\epsilon
  (T^h_{\epsilon+\omega'}{\cal L}_{i,j}^{++s}
  +T^h_{\epsilon+\omega}{\cal L}_{i,j}^{s+-}
  +T^h_{\epsilon}{\cal L}_{i,j}^{-s-})
\end{equation}
with
\begin{equation}
  {\cal L}_{2,j}^{s_1s_2s_3}=\frac{
  {\rm Tr}[\hat{\tau}_0
    -\hat{g}^{s_1}_{\epsilon+\omega}\hat{g}^{s_2}_{\epsilon}
    -\hat{\tau}_l\hat{g}^{s_1}_{\epsilon+\omega}\hat{\tau}_l
    \hat{g}^{s_3}_{\epsilon+\omega'}
    -\hat{\tau}_{l'}\hat{g}^{s_2}_{\epsilon}\hat{\tau}_{l'}
    \hat{g}^{s_3}_{\epsilon+\omega'}]}
{2(x+\zeta^{s_1}_{\epsilon+\omega}+\zeta^{s_3}_{\epsilon+\omega'})
    (x+\zeta^{s_2}_{\epsilon}+\zeta^{s_3}_{\epsilon+\omega'})}    
\end{equation}
and
\begin{equation}
  {\cal L}_{3,j}^{s_1s_2s_3}=\frac{
  {\rm Tr}[\hat{\tau_1}(\hat{\tau}_0
    -\hat{g}^{s_1}_{\epsilon+\omega}\hat{g}^{s_2}_{\epsilon}
    +\hat{\tau}_l\hat{g}^{s_1}_{\epsilon+\omega}\hat{\tau}_l
    \hat{g}^{s_3}_{\epsilon+\omega'}
    +\hat{\tau}_{l'}\hat{g}^{s_2}_{\epsilon}\hat{\tau}_{l'}
    \hat{g}^{s_3}_{\epsilon+\omega'})]}
{2(x+\zeta^{s_1}_{\epsilon+\omega}+\zeta^{s_3}_{\epsilon+\omega'})
    (x+\zeta^{s_2}_{\epsilon}+\zeta^{s_3}_{\epsilon+\omega'})}.    
\end{equation}
$(l,l')=(0,3)$ and $(3,0)$ for $j=1$ and $2$, respectively.
The result in the case of the normal state~\cite{schmidt} 
is obtained by setting $\Delta=0$ in the above equations.

\subsubsection{Maximally crossed term}

\begin{figure}
  \includegraphics[width=11.5cm]{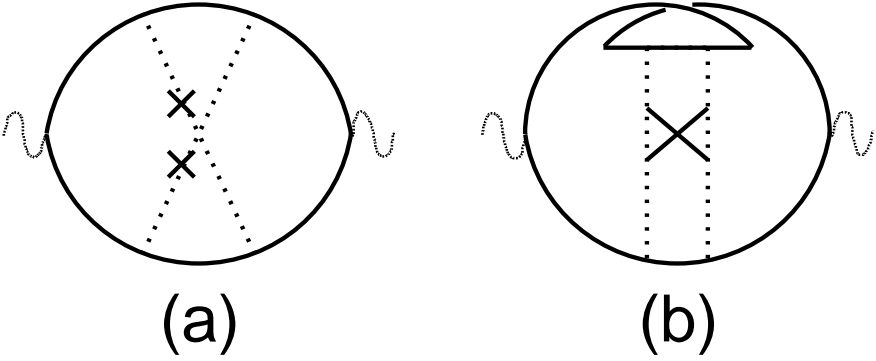}
  \caption{\label{fig:6}
    (a) Diagram of the conductivity corrected by
    the crossed impurity scattering at the lowest order.
    (b) Diagram of the conductivity
    corrected by the maximally crossed impurity scattering
    (the Cooperon term).
  }
\end{figure}

The maximally crossed (MC) term is the vertex correction, which does not
include the screened Coulomb interaction
and the superconducting fluctuation explicitly,
but gives a weak localization correction
to the conductivity by the coherent
backscattering effect.~\cite{gorkov,smith}
(The contribution of this term to the one-particle spectrum
vanishes and is not included in Figs. 1 and 2.)
The diagram of this term is shown in
Fig.~\ref{fig:6}(b).
As derived in Appendix A.4, the linear absorption by the maximally
crossed term is written as
\begin{equation}
  \frac{{\rm Re}\sigma^{MC}_{\omega}}{\sigma_0}
  =
  \frac{3\sqrt{3\tau}}{8\pi\omega(k_Fl)^2}
  \int d x \sqrt{x}
  \int d\epsilon(T^h_{\epsilon+\omega}-T^h_{\epsilon})
       {\rm Re}\sum_{s=\pm}s\frac{{\rm Tr}
         [\hat{\tau}_0+\hat{g}^+_{\epsilon+\omega}\hat{g}^s_{\epsilon}]}
       {2(x+\zeta_{\epsilon+\omega}^++\zeta_{\epsilon}^s)}.
       \label{eq:condMC}
\end{equation}

\section{Results}

\subsection{Numerical calculations}

In this section, we numerically evaluate 
Eqs. (\ref{eq:condMT0}), (\ref{eq:condMT}), (\ref{eq:condDOS0}),
(\ref{eq:condDOS}), (\ref{eq:condAL}), and (\ref{eq:condMC}).
The ranges of integrations over $x=D q^2$, $\epsilon$, and  $\omega'$
in these equation are $x<1/\tau$, $|\epsilon|$, and $|\omega'|<1/\tau$,
in which the approximation Eq. (\ref{eq:XtoDiff}) holds.
(The low-energy properties are given by this range.)
We make the variables symmetrical in advance, such as
$\epsilon_{1\textendash 4}$
below Eq. (\ref{eq:MT0cdot}). 
The superconducting gap at $T=0$
is taken to be the unit of energy
($\Delta_0=1$),
and the  electron--phonon coupling $p$ is determined
by the gap equation.
$c_q$ ($=\pi \omega_p^2\tau/2D q^2
\gg p$
with $\omega_p=\sqrt{4\pi n_e e^2/m}$ the plasma frequency)
in the denominator and the numerator of
$\Gamma_{3,2,4}(q)$ cancel out each other as in Eqs. (45)--(47)
in Ref. 25.
(The effective interaction becomes
independent of $e^2$.~\cite{altshuler})
The relation between $\alpha=1/2\tau$ and
$k_Fl$ is fixed to
$k_Fl/2\tau=E_F=300\Delta_0$, where $E_F$ is the Fermi energy.

The calculated results of $\Gamma_i(q)$
at $T=0$ are given in Ref. 25,
in which it is shown that
terms including the Coulomb interaction
($\Gamma_{3,2,4}$) 
are larger than other terms 
($\Gamma_{0,1}$).
In the case of the absorption spectrum,
it can also be shown that terms including the Coulomb interaction
predominantly contribute to
the correction term.
We will not show 
correction terms decomposed for each vertex ($\Gamma_i$)
explicitly below.

\begin{figure}
  \includegraphics[width=11.5cm]{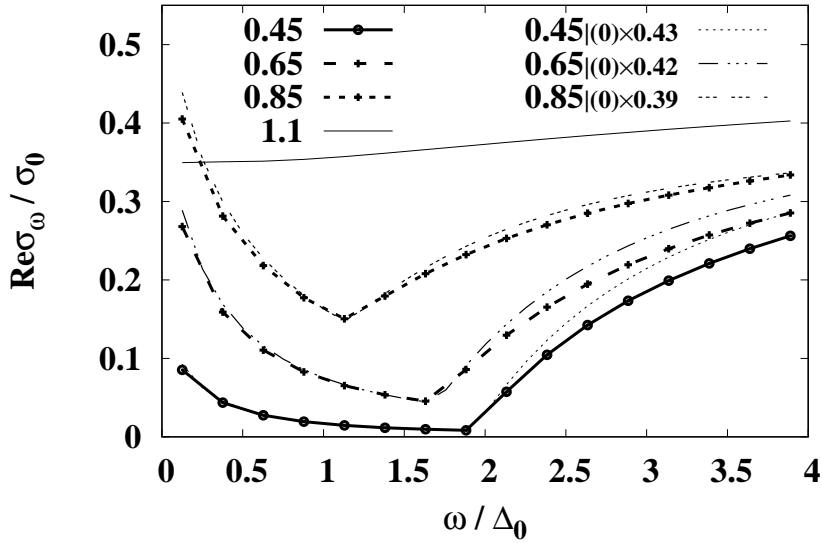}
  \caption{\label{fig:7}
    Dependence of ${\rm Re}\sigma_{\omega}/\sigma_0$ on $\omega$
    for $k_Fl=2.5$.
    The numerical values on the left of lines
    are the values of $T/T_c$. 
    The dotted thin lines represent
    $c_0\times{\rm Re}\sigma^{(0)}_{\omega}/\sigma_0$.
    $c_0$ is a constant with its value indicated
    as $(0)\times c_0$.
  }
\end{figure}

The dependence of 
${\rm Re}\sigma_{\omega}/\sigma_0$ on $\omega$
for several values of $T/T_c$ is shown in 
Fig.~\ref{fig:7}. 
Here,
$\sigma_{\omega}=
\sigma^{(0)}_{\omega}+\sigma^{\rm vc}_{\omega}$
with $\sigma^{\rm vc}_{\omega}
=\sigma^{MT0}_{\omega}+\sigma^{MT}_{\omega}
+\sigma^{DOS0}_{\omega}+\sigma^{DOS}_{\omega}
+\sigma^{AL}_{\omega}+\sigma^{MC}_{\omega}$
[${\rm Re}\sigma_{\omega}=
  {\rm Re}\sigma_{\omega}^{(0)}\left(
  1+{\rm Re}\sigma^{\rm vc}_{\omega}/
  {\rm Re}\sigma_{\omega}^{(0)}\right)$].
${\rm Re}\sigma^{(0)}_{\omega}$ is the
real part of the conductivity
given by the MB theory
(i.e., the linear absorption without vertex corrections)
and is written as
\begin{equation}
  \frac{{\rm Re}\sigma^{(0)}_{\omega}}{\sigma_0}
  =
  \frac{-1}{2\omega}
  \int d\epsilon(T^h_{\epsilon+\omega}-T^h_{\epsilon})
       {\rm Re}\sum_{s=\pm}s\frac{1}{2}{\rm Tr}
       [\hat{g}^+_{\epsilon+\omega}\hat{g}^s_{\epsilon}].
       \label{eq:mbcond}
\end{equation}
The comparison of
${\rm Re}\sigma_{\omega}/\sigma_0$
with the MB formula
${\rm Re}\sigma^{(0)}_{\omega}/\sigma_0$
is shown in Fig. 7. 
In this figure, $(0)\times c_0$
represents the result of 
$c_0\times {\rm Re}\sigma^{(0)}_{\omega}/\sigma_0$.
$c_0$ is chosen such that
$c_0\times {\rm Re}\sigma^{(0)}_{\omega}/\sigma_0$
overlaps
${\rm Re}\sigma_{\omega}/\sigma_0$
in the range of $\omega\lessapprox 2\Delta$.
This result shows that
${\rm Re}\sigma_{\omega}/\sigma_0$
is not proportional to the MB conductivity. 
The suppression of the conductivity
by vertex corrections
is larger
at $\omega>2\Delta$
than at $\omega<2\Delta$ for low temperatures.
As a result, the excitations below
$2\Delta$ (i.e., thermal excitations) seem to be relatively
enhanced as compared with the behavior of 
the MB conductivity.

\begin{figure}
  \includegraphics[width=11.cm]{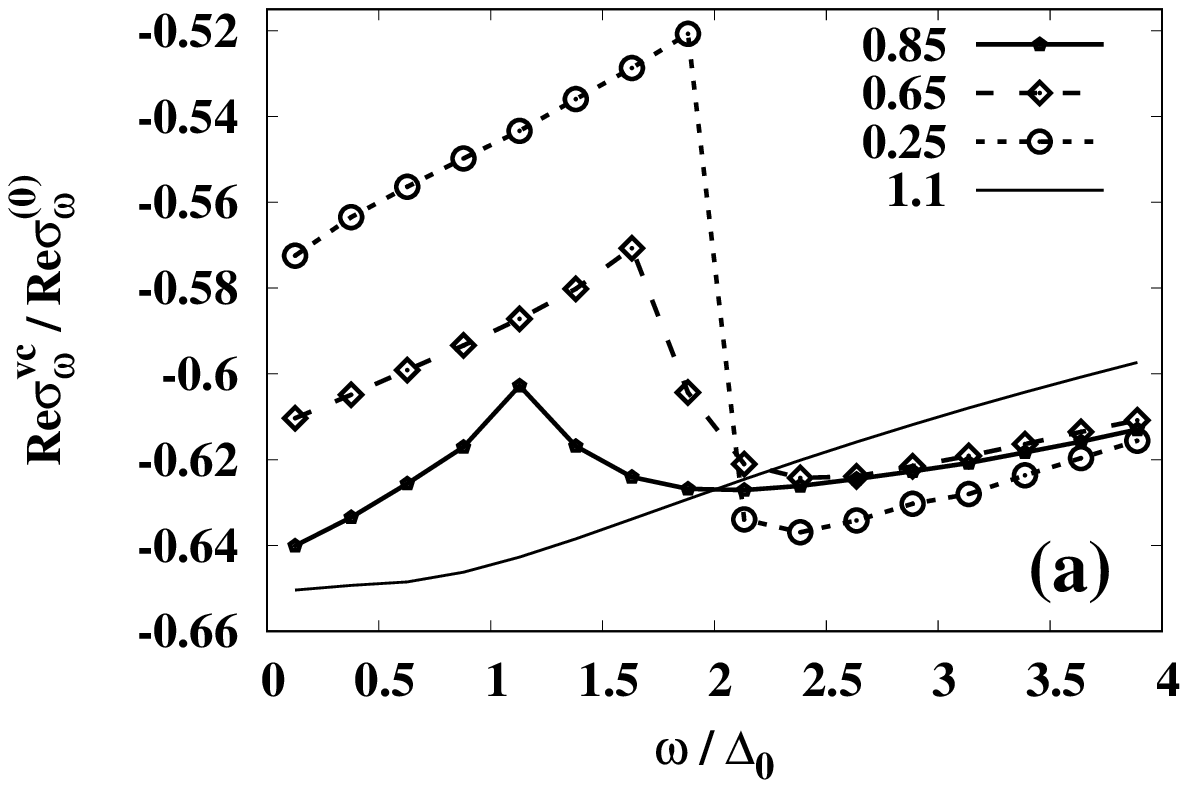}
  \includegraphics[width=11.cm]{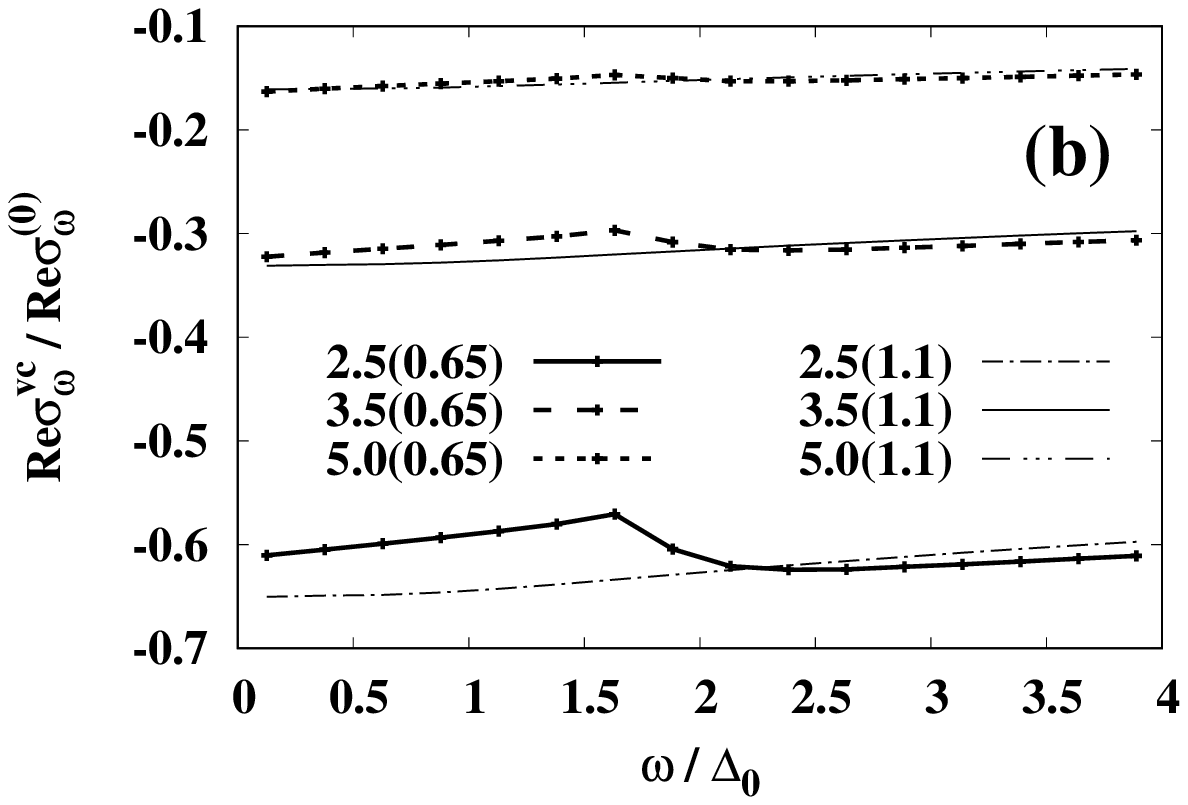}
  \caption{\label{fig:8}
    (a) Ratio of the vertex correction to the MB conductivity,
    ${\rm Re}\sigma^{\rm vc}_{\omega}/{\rm Re}\sigma^{(0)}_{\omega}$,
    for $k_Fl=2.5$.
    The numerical values on the left of lines
    represent the values of $T/T_c$. 
    (b) ${\rm Re}\sigma^{\rm vc}_{\omega}/{\rm Re}\sigma^{(0)}_{\omega}$
    for several values of $k_Fl$.
    The numerical values indicate the values of
    $k_Fl$ and $T/T_c$
    ($k_Fl=2.5$, $3.5$, and $5.0$, and $T/T_c=0.65$ and $1.1$).
  }
\end{figure}

The ratio of vertex corrections to
the MB conductivity,
${\rm Re}\sigma^{\rm vc}_{\omega}/{\rm Re}\sigma^{(0)}_{\omega}$,
is shown in 
Fig.~\ref{fig:8}. 
Figure 8(a) shows that the difference in
${\rm Re}\sigma^{\rm vc}_{\omega}/{\rm Re}\sigma^{(0)}_{\omega}$ 
between $\omega>2\Delta$ and $\omega<2\Delta$
becomes significant at low temperatures.
The suppression of 
${\rm Re}\sigma_{\omega}$ by vertex corrections
is larger (smaller) than that in the normal state
($T/T_c=1.1$) for $\omega>2\Delta$ ($\omega<2\Delta$).
The vertex correction is less effective for
the thermal excitation ($\omega<2\Delta$). 
Figure 8(b) shows that
the absolute value of 
${\rm Re}\sigma^{\rm vc}_{\omega}$ is roughly proportional to
$1/(k_Fl)^2$, as expected from the analytical expressions
in the previous section.
There also exists a difference between
$\omega>2\Delta$ and $\omega<2\Delta$
for other values of $k_Fl$, which is similar to 
the case of $k_Fl=2.5$.

\begin{figure}
  \includegraphics[width=8.5cm]{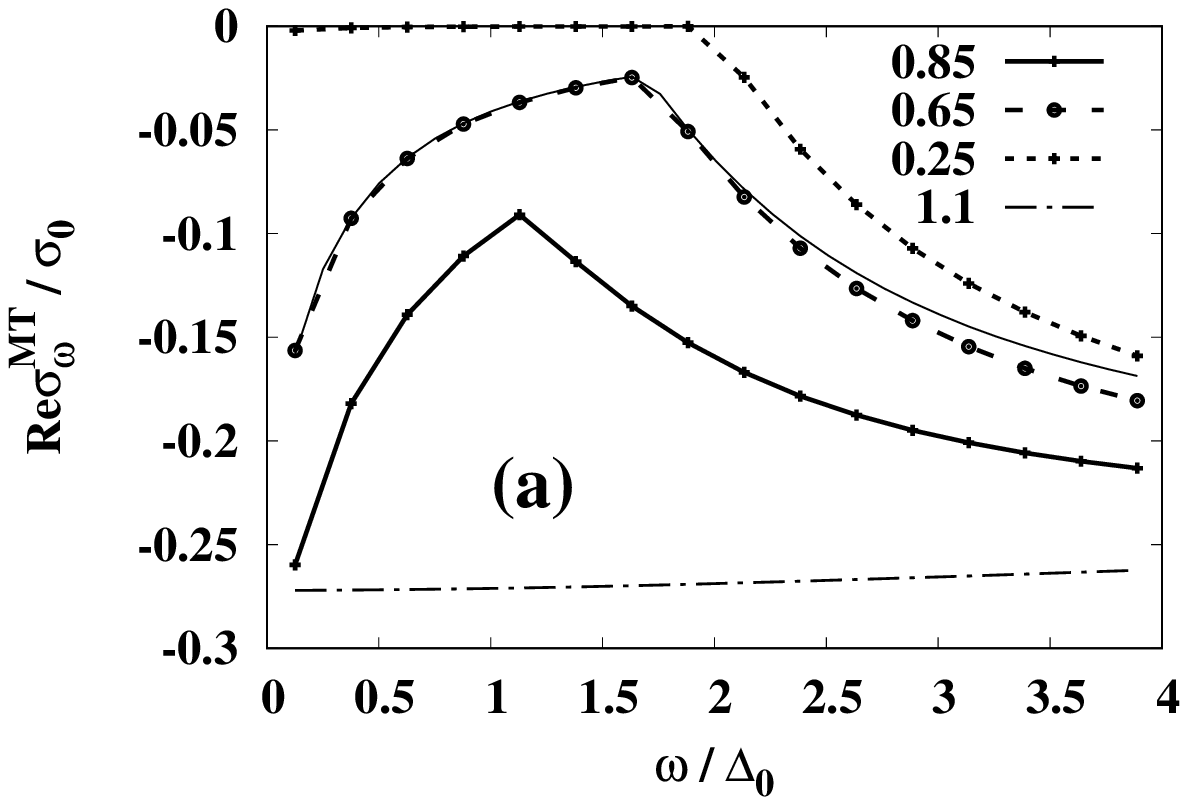}
  \includegraphics[width=8.5cm]{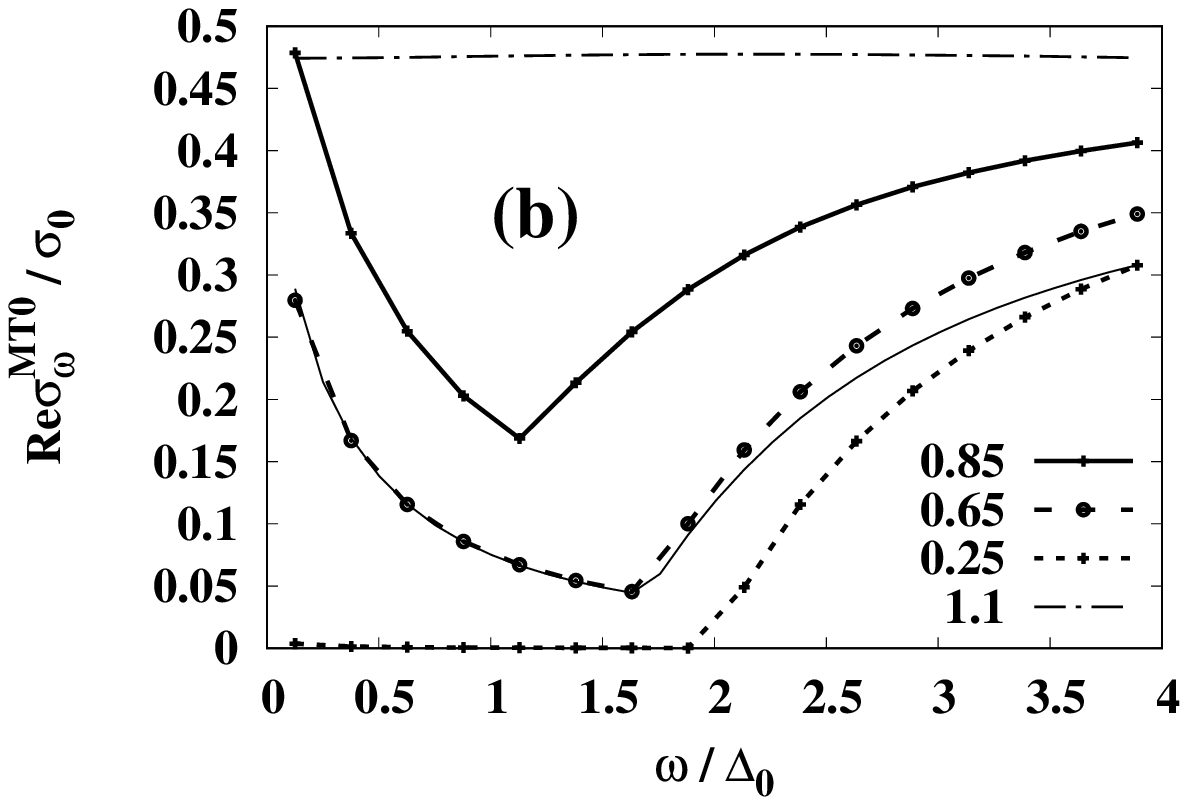}
  \includegraphics[width=8.5cm]{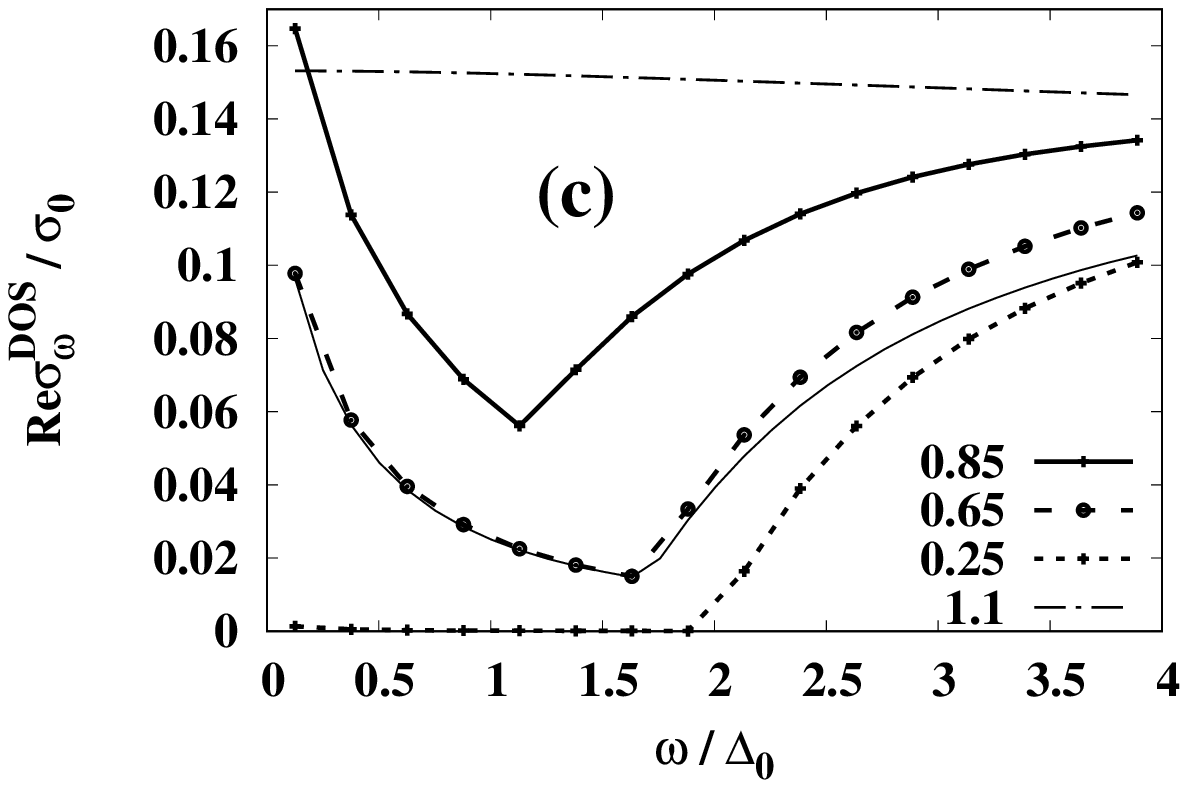}
  \includegraphics[width=8.5cm]{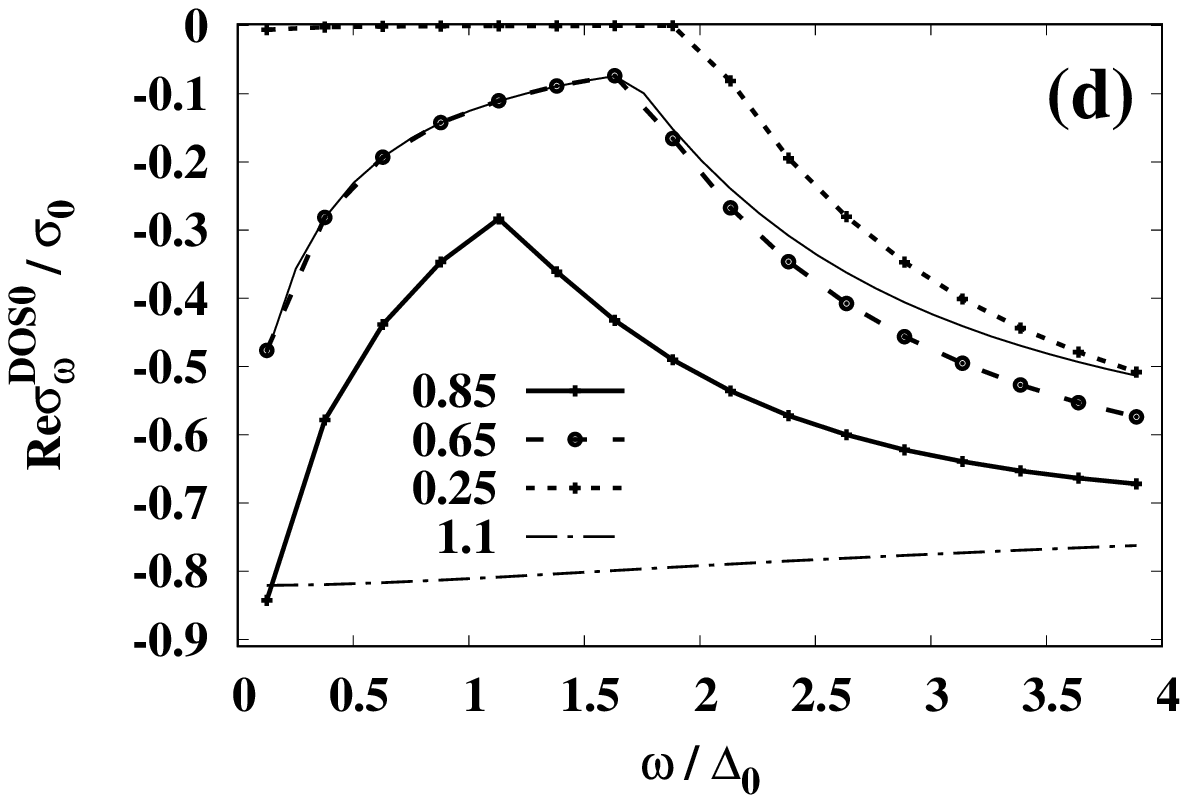}
  \includegraphics[width=8.5cm]{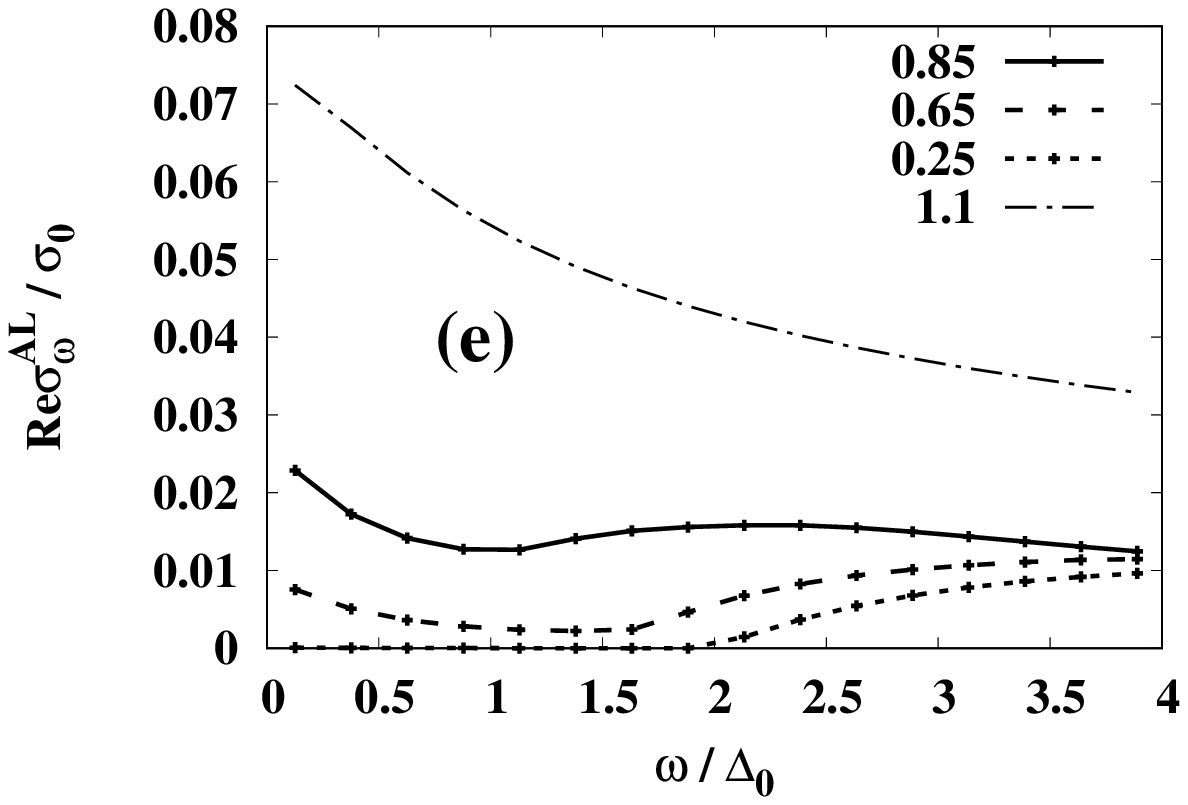}
  \includegraphics[width=8.5cm]{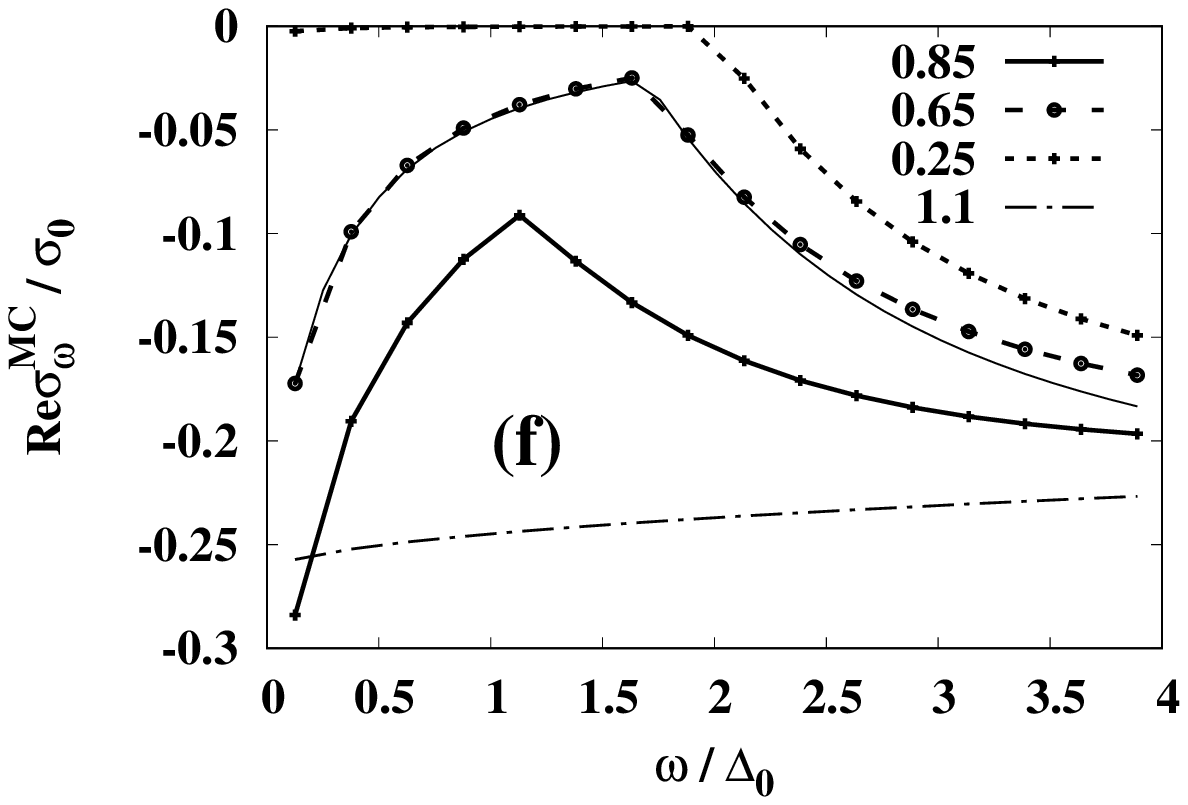}
  \caption{\label{fig:9}
    Components of
    ${\rm Re}\sigma^{\rm vc}_{\omega}/\sigma_0$ for $k_Fl=2.5$.
    The numerical values on the left of lines are the values of
    $T/T_c$.
    The thin solid lines show the Mattis--Bardeen formula
    at $T/T_c=0.65$
    with a constant value multiplied
    ($c_0\times {\rm Re}\sigma^{(0)}_{\omega}/\sigma_0$).
    (a) ${\rm Re}\sigma^{MT}_{\omega}/\sigma_0$.
    $c_0=-0.23$.
    (b) ${\rm Re}\sigma^{MT0}_{\omega}/\sigma_0$.
    $c_0=0.42$.
    (c) ${\rm Re}\sigma^{DOS}_{\omega}/\sigma_0$.
    $c_0=0.7$.
    (d) ${\rm Re}\sigma^{DOS0}_{\omega}/\sigma_0$.
    $c_0=-0.4$.
    (e) ${\rm Re}\sigma^{AL}_{\omega}/\sigma_0$.
    (f) ${\rm Re}\sigma^{MC}_{\omega}/\sigma_0$.
    $c_0=-0.25$.
  }
\end{figure}

The components of 
${\rm Re}\sigma^{\rm vc}_{\omega}/\sigma_0$
are shown in 
Fig.~\ref{fig:9}. 
Figure 9 shows that 
${\rm Re}\sigma^{MT,DOS0,MC}_{\omega}/\sigma_0$
(${\rm Re}\sigma^{MT0,DOS,AL}_{\omega}/\sigma_0$)
takes a negative (positive) value.
These vertex corrections are seemingly proportional
to the MB conductivity.
The comparison between
vertex corrections and 
$c_0\times {\rm Re}\sigma^{(0)}_{\omega}/\sigma_0$ 
shows that this is not the case.
Here, $c_0$ is chosen in the same way as in Fig. 7.
In the cases of ${\rm Re}\sigma^{MT,MT0,DOS,DOS0}_{\omega}$,
the deviations from the MB conductivity
show a similar $\omega$-dependence with each other.
When
we choose $c_0$ so that
$c_0\times {\rm Re}\sigma^{(0)}_{\omega}/\sigma_0$
overlaps ${\rm Re}\sigma^{MT,MT0,DOS,DOS0}_{\omega}$
around $\omega\lessapprox 2\Delta$,
each of $|{\rm Re}\sigma^{MT,MT0,DOS,DOS0}_{\omega}|$
takes values larger than
$|c_0\times {\rm Re}\sigma^{(0)}_{\omega}/\sigma_0|$
for $\omega>2\Delta$.
This indicates that
these four terms give 
larger corrections 
in ${\rm Re}\sigma_{\omega}$ for $\omega>2\Delta$
than for $\omega<2\Delta$.
The maximally crossed term shows 
an opposite tendency.
The correction by the thermal excitation
is larger than that by the excitation above $2\Delta$
in the case of $|{\rm Re}\sigma^{MC}_{\omega}|$.
The AL term is smaller than other terms
and negligible in the superconducting state,
although ${\rm Re}\sigma^{AL}_{\omega}$
in the normal state
is enhanced for small $\omega$ by 
the superconducting fluctuation near $T_c$.
This smallness of the AL term originates
from the fact that one of two 
fluctuation modes is the amplitude mode
($\Gamma_1$), 
as shown in Eq. (\ref{eq:QAL}).
Note that $\Gamma_1$ does not include the Coulomb interaction effect.

\begin{figure}
  \includegraphics[width=11.5cm]{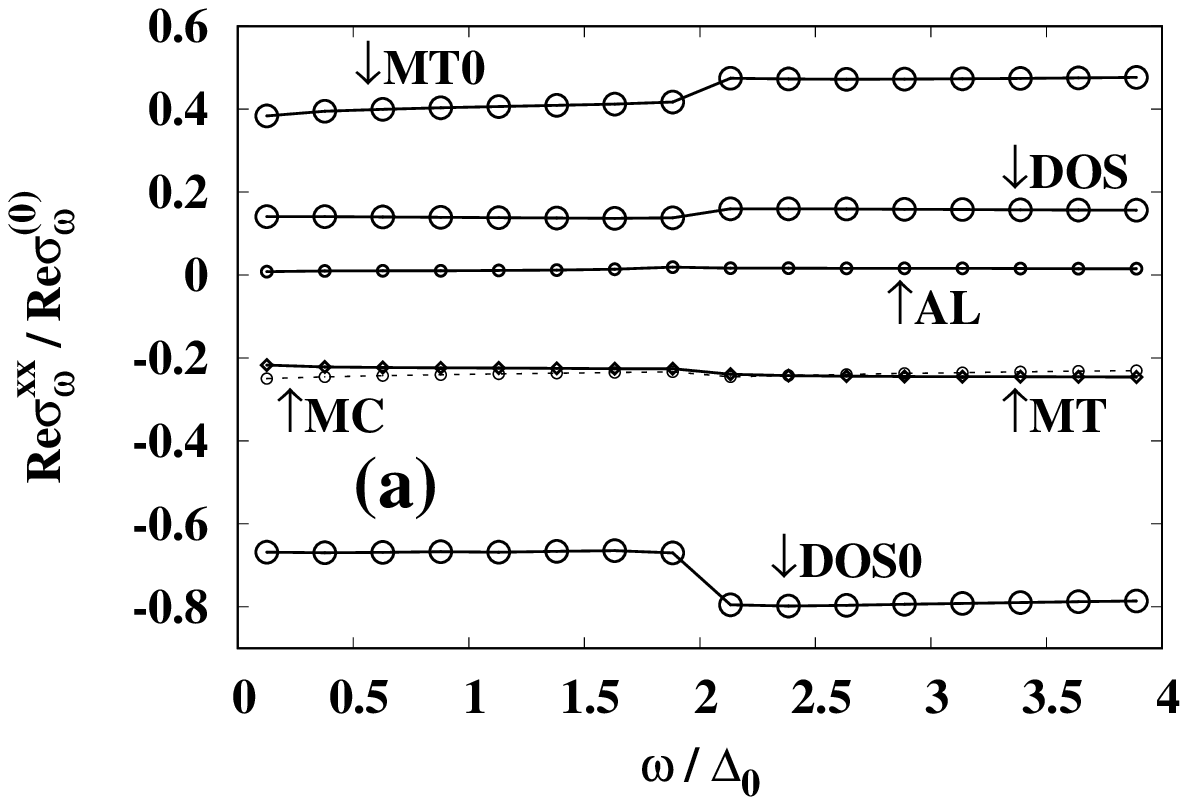}
  \includegraphics[width=11.5cm]{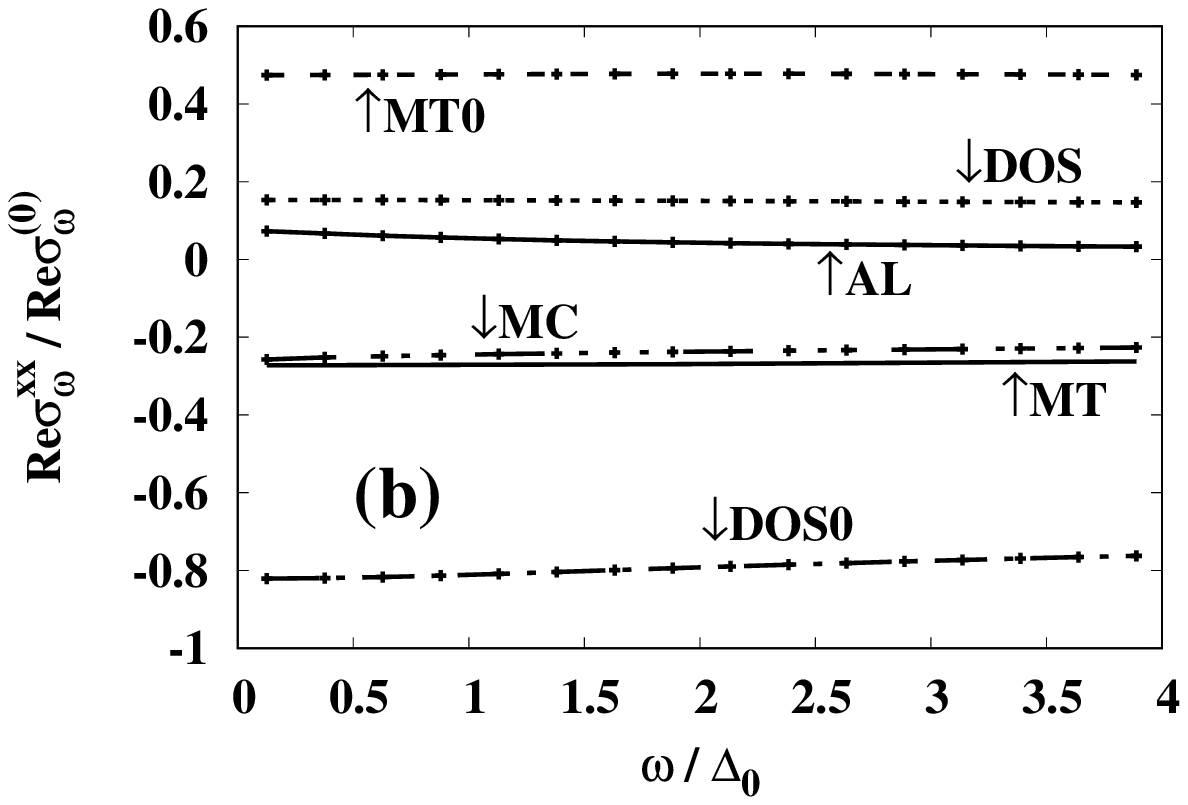}
  \caption{\label{fig:10}
Ratio 
${\rm Re}\sigma^{\rm xx}_{\omega}/{\rm Re}\sigma^{(0)}_{\omega}$
for $k_Fl=2.5$.
Here, ${\rm xx}=MT0$, $MT$, $DOS0$, $DOS$, $AL$, or $MC$
as indicated in the figure.
(a) $T/T_c=0.45$ in the superconducting state. 
(b) $T/T_c=1.1$ in the normal state. 
  }
\end{figure}

The ratios of components of vertex corrections
to the MB conductivity
are shown in 
Fig.~\ref{fig:10}. 
The relationship between the sizes of vertex
corrections in the superconducting state
is similar to that in the normal state.
For $T/T_c=0.45$, the absolute values of the MT and DOS terms 
are smaller for $\omega<2\Delta$ 
than for $\omega>2\Delta$,
and show a variation around $\omega\simeq 2\Delta$.
The comparison of the absolute values between
vertex corrections shows that 
${\rm Re}\sigma^{DOS0}_{\omega}$
gives a predominant contribution in 
${\rm Re}\sigma_{\omega}$, although there is a cancellation
among vertex corrections.
Thus, the difference in
${\rm Re}\sigma^{DOS0}_{\omega}$
between $\omega<2\Delta$ and $\omega>2\Delta$
mainly leads to the deviation
of ${\rm Re}\sigma_{\omega}$ from the MB-like behavior.
For $T/T_c=1.1$, 
${\rm Re}\sigma^{DOS0}_{\omega}$
is suppressed at small values of $\omega$,
which causes a suppression of 
${\rm Re}\sigma_{\omega}$ at low frequencies,
as shown in Fig. 7.

\subsection{Origin of the variation of
  the correction term around $\omega=2\Delta$}
  
\begin{figure}
  \includegraphics[width=11.5cm]{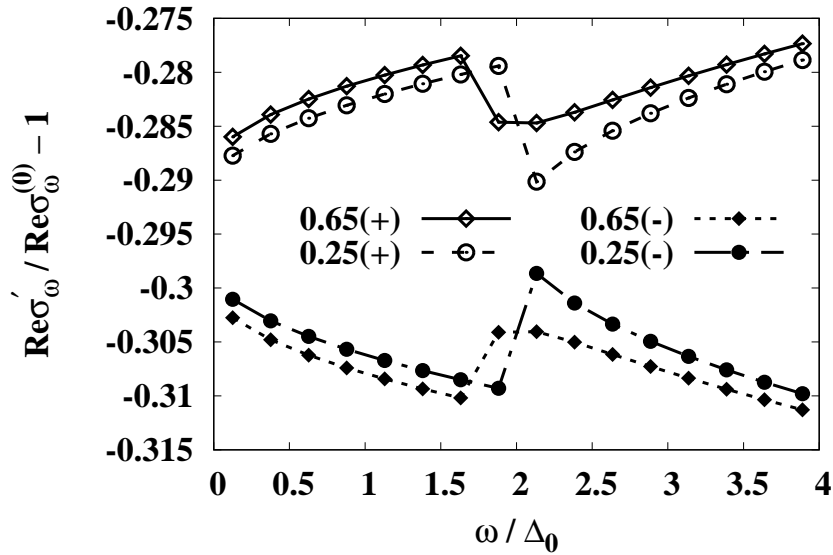}
  \caption{\label{fig:11}
    Correction part of the conductivity, 
    ${\rm Re}\sigma^{'}_{\omega}/{\rm Re}\sigma^{(0)}_{\omega}-1$,
    induced by the suppression of the density of states.
    The numerical values indicate temperatures ($T/T_c$).
    The signs $(\pm)$ correspond to $s'=\pm$ in the main text.
  }
\end{figure}

In this subsection, we show that
the variation of the ratio
${\rm Re}\sigma^{\rm vc}_{\omega}/{\rm Re}\sigma^{(0)}_{\omega}$
around $\omega\simeq 2\Delta$ is related to
the correction to the one-particle spectrum.
Equation (\ref{eq:mbcond})
is rewritten as 
${\rm Re}\sigma^{(0)}_{\omega}/\sigma_0=
({\rm Re}\sigma^{(0)a}_{\omega}
+{\rm Re}\sigma^{(0)b}_{\omega})/\sigma_0$
with 
${\rm Re}\sigma^{(0)a}_{\omega}/\sigma_0  =
\int_{\Delta+\omega/2}^{\infty} d\epsilon
h_{\epsilon,\omega}$ and
${\rm Re}\sigma^{(0)b}_{\omega}/\sigma_0  =
-\int_{0}^{\omega/2-\Delta} d\epsilon
\theta(\omega-2\Delta)h_{\epsilon,\omega}$
for $\omega>0$.
Here, $h_{\epsilon,\omega}=
(T^h_{\epsilon+\omega/2}-T^h_{\epsilon-\omega/2}){\rm Tr}
[{\rm Im}\hat{g}^+_{\epsilon+\omega/2}
  {\rm Im}\hat{g}^+_{\epsilon-\omega/2}]/\omega$.
${\rm Re}\sigma_{\omega}^{(0)a}$ is a
decreasing function with respect to $\omega$ and
${\rm Re}\sigma_{\omega}^{(0)b}$ takes finite values
only for $\omega>2\Delta$.
The weak localization effect on the one-particle spectrum
is approximately taken into account as
${\rm Im}\hat{g}^+_{\epsilon}\times
[1-(1-s'\sqrt{(|\epsilon|-\Delta)\tau})/(k_Fl)^2]=:
{\rm Im}\hat{g'}^+_{\epsilon}$.
$s'=+$ $(-)$ means that
the suppression of the density of states is
large close to (apart from) the gap edge.
Here, the case of $s'=-$ is used for comparison
and does not correspond to the real system.
We calculate the conductivity including this effect
(${\rm Re}\sigma'_{\omega}=
{\rm Re}\sigma^{'a}_{\omega}
+{\rm Re}\sigma^{'b}_{\omega}$)
by replacing $\hat{g}^+_{\epsilon}$
by $\hat{g'}^+_{\epsilon}$ in the above $h_{\epsilon,\omega}$. 
The correction part of the conductivity
${\rm Re}\sigma'_{\omega}/{\rm Re}\sigma^{(0)}_{\omega}-1$
is shown in Fig.~\ref{fig:11}. 
The large variation of
${\rm Re}\sigma^{'}_{\omega}/{\rm Re}\sigma^{(0)}_{\omega}-1$
occurs around $\omega\simeq 2\Delta$
with the same (opposite) tendency for $s'=+$ $(-)$ as
that of
${\rm Re}\sigma^{\rm vc}_{\omega}/{\rm Re}\sigma^{(0)}_{\omega}$.

In ${\rm Re}\sigma^{'a}_{\omega}$, 
which is given by the thermal excitations,
$\epsilon+\omega/2\simeq 3\Delta$
and $\epsilon-\omega/2\simeq \Delta$
for $\omega\simeq 2\Delta$
(the contribution from $\epsilon\simeq \Delta+\omega/2$
is large in the integration
at low temperatures), 
and then only ${\rm Im}\hat{g'}^+_{\epsilon-\omega/2}$
is put on the gap edge.
On the other hand, 
in ${\rm Re}\sigma^{'b}_{\omega}$, 
which is given by excitations across the gap with $\omega\gtrsim 2\Delta$, 
$\epsilon+\omega/2\simeq \Delta$
and $\epsilon-\omega/2\simeq -\Delta$
for $\omega\simeq 2\Delta$, and
then both ${\rm Im}\hat{g'}^+_{\epsilon\pm\omega/2}$
are put on the gap edge.
Thus, ${\rm Re}\sigma^{'b}_{\omega}$ is
more sensitive to the values of the density
of states near the gap edge than
${\rm Re}\sigma^{'a}_{\omega}$
in the case of $\omega\simeq 2\Delta$.
When the suppression of the density of states
is larger near the gap edge than
apart from the gap edge (corresponding to the case of $s'=+$),
the correction part of 
$|{\rm Re}\sigma^{'b}_{\omega}|$ for $\omega\gtrsim 2\Delta$ is
larger than that of 
$|{\rm Re}\sigma^{'a}_{\omega}|$ for $\omega\lesssim 2\Delta$. 
This leads to the negative slope of
${\rm Re}\sigma'_{\omega}/{\rm Re}\sigma^{(0)}_{\omega}-1$
around $\omega\simeq 2\Delta$, as shown in Fig. 11.

\section{Summary and Discussion}

In this study, we calculated the weak localization correction to
the linear absorption of three-dimensional s-wave superconductors.
This correction is caused by 
the coherent backscattering 
and the interaction effect (Coulomb interaction
and superconducting fluctuations) enhanced
by the diffusion term.
Numerical calculations show that
the correction effect is greater on
the excitations across the gap ($\omega>2\Delta$) 
than on the thermal excitations
($\omega<2\Delta$).
This tendency is shown to be related to
the weak localization correction to the one-particle spectrum.
The latter is known as
the Altshuler--Aronov effect, which was confirmed in 
experiments~\cite{sacepe,chand}.

We showed that the weak localization effect can be made
clear by taking the ratio of the correction term to
the MB formula.
The dependence of this ratio on the frequency
shows a large variation near $\omega\simeq 2\Delta$.
This variation indicates that
the correction term
relatively enhances the absorption by the thermal excitations 
as compared with that by the excitations across the gap
and blurs the absorption edge in the spectrum.
The experimental result
in a strongly disordered system 
shows a qualitatively similar tendency.~\cite{cheng}
The quantitative result is not explained
by our calculation because this system is
outside the range of the weak localization.

Even in the case of less disordered systems,
by taking the ratio of the absorption spectrum
to the MB formula,
the weak localization effect should be visible
as a variation of this ratio around $\omega\simeq 2\Delta$.
We consider that this ratio
can be used as a sign of the weak localization.

In order to deal with a more disordered system
(i.e., outside the weak localization regime)
such as some experiments,~\cite{cheng,simmendinger,pracht}
the following should be considered
theoretically as future issues.
Since the calculations in this paper are based on the perturbation
theory, one way to improve the calculation
is to take account of higher-order terms.
This is needed when we consider the metal-insulator transition.
As another possibility, it is conceivable to
change the treatment of the impurity scattering,
for example, by considering an inhomogeneous case.~\cite{larkin,ghosal}
In this study, we considered a homogeneously disordered
system, but it is meaningful to see how 
the Coulomb interaction (a predominant factor in
the weak localization effect) is incorporated
in the case of inhomogeneous systems.~\cite{seibold}

\section*{Acknowledgment}

The numerical computation in this work was carried out
at the Yukawa Institute Computer Facility.

  \appendix
  \section{Derivation of Correction Terms
  for Linear Absorption}

In this Appendix, we give a detailed derivation of
the results shown in Sect. 3.2.
We obtain
expressions of correction terms 
Eqs. (\ref{eq:condMT0}), (\ref{eq:condMT}), (\ref{eq:condDOS0}),
(\ref{eq:condDOS}), (\ref{eq:condAL}), and (\ref{eq:condMC})
by calculating Eqs. (\ref{eq:JdSdA})--(\ref{eq:actionSd})
with the use of Eqs. (\ref{eq:G4by4})--(\ref{eq:Vi4by4}).
The calculation is performed by expanding
$e^{i S'}$ in accordance with Figs. 3--6,
and we incorporate the effects of interactions
(i.e., the screened Coulomb interaction, superconducting fluctuation,
and scattering by impurities) in accordance with Fig. 1.
In the following subsections, we show the equations obtained
by this process.

\subsection{Maki--Thompson term}

The real part of the conductivity shown in Fig. 3(a)
is written as
\begin{equation}
  \begin{split}
  {\rm Re}\sigma^{MT0}_{\omega}=&
    \frac{e^2}{2\omega N^3}\sum_{\mib q'}\int\frac{d\omega'}{2\pi i}
  \int\frac{d\epsilon}{2\pi i}{\rm Im}
  \Bigl\{ 
  \left(\frac{\pi \rho_0}{2}\right)^{-1} \\
  &\times \sum_{i=0,1,2,3,4}
  \left(\sum_{{\cal S}_1}
  C^t_{\omega'}[\Gamma_i(q')-\Gamma^*_i(q')]
  +\sum_{{\cal S}_2}\Gamma_i(q')
  +\sum_{{\cal S}_3}\Gamma^*_i(q')\right)\\&\times
  \frac{1}{N^3}\sum_{\mib k}
  {\rm Tr}[\hat{I}^{s_4,b'}_{\epsilon+\omega'}
\hat{G}_{k+q'}^{s_4}v_{k+q'}\hat{G}_{k+q+q'}^{s_1}
    \hat{I}^{s_1,a}_{\epsilon+\omega+\omega'}
    \hat{Y}_{i}\hat{I}^{s_2,a'}_{\epsilon+\omega}
    \hat{G}_{k+q}^{s_2}v_{k}\hat{G}_{k}^{s_3}
        \hat{I}^{s_3,b}_{\epsilon}\hat{Y}'_{i}]
  \Bigr\}.
  \end{split}
  \label{eq:MT0tr}
\end{equation}
$\Gamma_{i=0,1,2,3,4}$ indicate the interaction effects 
[Eqs. (\ref{eq:Gam32d})--(\ref{eq:Gam1})],
and the derivation of these equations
is given in Ref. 25.
$\sum_{{\cal S}_{1,2,3}}$ indicate
the summations taken over $(s_1,s_2,s_3,s_4)$ with
\begin{equation}
(s_1,s_2,s_3,s_4)=\left\{
  \begin{matrix}
    (+,+,+,K), (+,+,K,-), (+,K,-,-), (K,-,-,-) & \text{for } {\cal S}_1
    \\
    (+,+,K, K), (+,K,-,K),(+,+,+,-), (-,-,-,+) & \text{for } {\cal S}_2
    \\
    (K,+,K,-), (K, K,-,-), (+,-,-,-), (-,+,+,+) & \text{for } {\cal S}_3.
     \end{matrix}
  \right.
\end{equation}
$(\hat{I})(\hat{I}')$
represents the vertex correction by the impurity scattering,
and it is written as 
\begin{equation}
  (\hat{I}^{s_1,x}_{\epsilon})(\hat{I}^{s_2,x'}_{\epsilon-\omega'})
  =(\hat{\tau}_0)(\hat{\tau}_0)+
  \frac{n_i u^2}{N^3}\sum_{\mib k}
  (\hat{\tau}_3\hat{G}_k^{s_1})(\hat{G}_{k-q'}^{s_2}\hat{\tau}_3)
  +\left(\frac{n_i u^2}{N^3}\right)^2\sum_{\mib k,\mib k'}
  (\hat{\tau}_3\hat{G}_k^{s_1}\hat{\tau}_3\hat{G}_{k'}^{s_1})
  (\hat{G}_{k'-q'}^{s_2}\hat{\tau}_3\hat{G}_{k-q'}^{s_2}\hat{\tau}_3)
  +\cdots
\end{equation}
($x=a,b$).
$(\hat{\;\;})(\hat{\;\;})$
represents a combination of matrices
and does not mean a product of matrices
  [$(\hat{A})(\hat{B})\ne \hat{A}\hat{B}$].
The summation taken over $\mib k$ results in
the following expression:
\begin{equation}
  (\hat{I}^{s_1,x}_{\epsilon})(\hat{I}^{s_2,x'}_{\epsilon-\omega'})
  =(\hat{\tau}_0)(\hat{\tau}_0)+
  \frac{X^{s_1,s_2}_{\epsilon,\epsilon-\omega'}
  }       {1-2X^{s_1,s_2}_{\epsilon,\epsilon-\omega'}}
       [(\hat{\tau}_3\hat{g}_{\epsilon}^{s_1})
    (\hat{g}_{\epsilon-\omega'}^{s_2}\hat{\tau}_3)
    +(\hat{\tau}_0)(\hat{\tau}_0)]
  \label{eq:MT0diff}
\end{equation}
with
\begin{equation}
  X^{s_1,s_2}_{\epsilon,\epsilon-\omega'}=
  \alpha\int_{-1}^{1}\frac{d({\rm cos}\theta)}{2}
  \frac{2\alpha+\zeta_{\epsilon}^{s_1}+\zeta_{\epsilon-\omega'}^{s_2}}
  {(2\alpha+\zeta_{\epsilon}^{s_1}+\zeta_{\epsilon-\omega'}^{s_2})^2
    +(v_F q'{\rm cos}\theta)^2}. 
\end{equation}
This expression describes the diffusion at a low energy:
\begin{equation}
    \frac{X^{s_1,s_2}_{\epsilon,\epsilon-\omega'}/\alpha
    }       {1-2X^{s_1,s_2}_{\epsilon,\epsilon-\omega'}}
    \simeq
    \frac{1}{D q'^2+\zeta^{s_1}_{\epsilon}+\zeta^{s_2}_{\epsilon-\omega'}}.
\label{eq:XtoDiff}
\end{equation}
$(\hat{Y}_{i})(\hat{Y}'_{i})$ indicates the vertices
of interactions $\Gamma_i$.
$(\hat{Y}_{i})(\hat{Y}'_{i})
=(\hat{\tau}_{i})(\hat{\tau}_{i})$
for $i=0,1,2,3$,
and
$(\hat{Y}_{i})(\hat{Y}'_{i})
=(\hat{\tau}_{3})(-i\hat{\tau}_{2})
+(i\hat{\tau}_{2})(\hat{\tau}_{3})$
for $i=4$.

In the dirty limit ($\Delta\tau\ll 1$), the following approximation
holds for the summation over ${\mib k}$:
\begin{equation}
  \begin{split}
&  \frac{1}{N^3}\sum_{\mib k}
  (\hat{G}_{k+q'}^{s_4}v_{k+q'}\hat{G}_{k+q+q'}^{s_1})
  (\hat{G}_{k+q}^{s_2}v_{k}\hat{G}_{k}^{s_3}) 
  \simeq 
  \frac{\pi \rho_0\tau^3}{4}
  \int_{FS}v_k^2
  \{
 4(\hat{g}^{s_4}_{\epsilon+\omega'}\hat{g}^{s_1}_{\epsilon+\omega+\omega'})
 (\hat{g}^{s_2}_{\epsilon+\omega}\hat{g}^{s_3}_{\epsilon})\\
& +
[ (\hat{g}^{s_4}_{\epsilon+\omega'}\hat{g}^{s_1}_{\epsilon+\omega+\omega'})
  +(\hat{\tau}_0)]
[ (\hat{g}^{s_2}_{\epsilon+\omega}\hat{g}^{s_3}_{\epsilon})
  + (\hat{\tau}_0)] 
+
[(\hat{g}^{s_4}_{\epsilon+\omega'}\hat{\tau}_3)
  +(\hat{\tau}_3\hat{g}^{s_1}_{\epsilon+\omega+\omega'})]
[(\hat{g}^{s_2}_{\epsilon+\omega}\hat{\tau}_3)
  +(\hat{\tau}_3\hat{g}^{s_3}_{\epsilon})] \},
  \end{split}
  \label{eq:MT0g4}
\end{equation}
where $\int_{FS}$ is the integration over the Fermi surface.
Using the above expressions,
we obtain the result for the MT0 term [Eq. (\ref{eq:condMT0})].

The real part of the conductivity shown in Fig. 3(b)
is given by Eq. (\ref{eq:MT0tr}) with
$(1/N^3)\sum_{\mib k}{\rm Tr}[\;\cdot\;]$ replaced by 
\begin{equation}
  \begin{split}
&  n_i u^2\left(\frac{1}{N^3}\right)^2\sum_{\mib k,\mib k'}
    {\rm Tr}[
\\
&      \hat{I}^{s_4,b'}_{\epsilon+\omega'}\hat{G}_{k'+q'}^{s_4}
    v_{k'+q'}\hat{G}_{k'+q+q'}^{s_1}
    \hat{I}^{s_1,c}_{\epsilon+\omega+\omega'}
    \hat{\tau}_3\hat{G}^{s_1}_{k+q+q'}
    \hat{I}^{s_1,a}_{\epsilon+\omega+\omega'}
    \hat{Y}_{i}\hat{I}^{s_2,a'}_{\epsilon+\omega} 
    \hat{G}^{s_2}_{k+q}v_{k}\hat{G}^{s_3}_{k}
    \hat{\tau}_3\hat{I}^{s_3,c'}_{\epsilon}\hat{G}^{s_3}_{k'}
    \hat{I}^{s_3,b}_{\epsilon}\hat{Y}'_{i}\\
&    +
 \hat{I}^{s_4,c}_{\epsilon+\omega'}\hat{\tau}_3
    \hat{G}_{k'+q'}^{s_4}
    v_{k'+q'}\hat{G}_{k'+q+q'}^{s_1}
        \hat{I}^{s_1,a}_{\epsilon+\omega+\omega'}
        \hat{Y}_{i}\hat{I}^{s_2,a'}_{\epsilon+\omega}
            \hat{G}^{s_2}_{k'+q}\hat{\tau}_3\hat{I}^{s_2,c'}_{\epsilon+\omega}
    \hat{G}^{s_2}_{k+q}v_{k}\hat{G}^{s_3}_{k}
        \hat{I}^{s_3,b}_{\epsilon}\hat{Y}'_{i}
        \hat{I}^{s_4,b'}_{\epsilon+\omega'}
        \hat{G}^{s_4}_{k+q'}
  ].
  \end{split}
\end{equation}
Using the same approximation as in Ref. 34
we obtain 
\begin{equation}
  \begin{split}
  \frac{1}{N^3}\sum_{\mib k}
(\hat{G}_{k+q'}^{s_4}
v_{k+q'}\hat{G}_{k+q+q'}^{s_1})
(\hat{G}^{s_3}_{k})
\simeq &
\frac{\pi\rho_0\tau^3}{2}
\int_{FS}v_k(\mib v_k\cdot \mib q')
[2(\hat{g}^{s_4}_{\epsilon+\omega'}\hat{g}^{s_1}_{\epsilon+\omega+\omega'})
(\hat{\tau}_3)
-(\hat{g}^{s_4}_{\epsilon+\omega'}\hat{\tau}_3)
(\hat{g}^{s_3}_{\epsilon}) \\&
-(\hat{\tau}_3\hat{g}^{s_1}_{\epsilon+\omega+\omega'})
(\hat{g}^{s_3}_{\epsilon})],
  \end{split}
  \label{eq:MTapprox1}
\end{equation}
and
$(1/N^3)\sum_{\mib k}
(\hat{G}^{s_1}_{k+q+q'})
(\hat{G}^{s_2}_{k+q}v_{k}\hat{G}^{s_3}_{k})$
is calculated in the same way.
Using these expressions and
\begin{equation}
  \frac{e^2}{2\alpha^2N^3}\sum_{\mib q}
  \left(\int_{FS}v_k({\mib v}_k\cdot {\mib q})\right)^2
  =
  \frac{\sigma_0\sqrt{3\tau}}{2(k_Fl)^2}\int d(D q^2)
  (D q^2)^{3/2},
\end{equation}
we obtain the result of the MT term with an additional diffuson
[Eq. (\ref{eq:condMT})].

\subsection{Density of states term}

The expression of the self-energy correction
term [Fig. 4(a)] is given by 
\begin{equation}
  \begin{split}
  {\rm Re}\sigma^{DOS0}_{\omega}=&
    \frac{e^2}{\omega N^3}\sum_{\mib q'}\int\frac{d\omega'}{2\pi i}
  \int\frac{d\epsilon}{2\pi i}{\rm Im}
  \Bigl\{ 
  \left(\frac{\pi \rho_0}{2}\right)^{-1}\\&
  \times\sum_{i=0,1,2,3,4}
  \left(\sum_{{\cal S}_1}
  C^t_{\omega'}[\Gamma_i(q')-\Gamma^*_i(q')]
  +\sum_{{\cal S}_2}\Gamma_i(q')
  +\sum_{{\cal S}_3}\Gamma^*_i(q')\right) \\
  &
\times    \frac{1}{N^3}\sum_{\mib k}
    {\rm Tr}[\hat{I}^{s_3,b'}_{\epsilon}
      \hat{G}_k^{s_3}v_{k}
      \hat{G}_{k-q}^{s_4}v_{k}\hat{G}_k^{s_1}
    \hat{I}^{s_1,a}_{\epsilon}
    \hat{Y}_{i}\hat{I}^{s_2,a'}_{\epsilon-\omega'} 
    \hat{G}_{k-q'}^{s_2}\hat{I}^{s_2,b}_{\epsilon-\omega'}\hat{Y}'_{i}]
    \Bigr\}
  \end{split}
  \label{eq:DOS0tr}
\end{equation}
[two diagrams in Fig. 4(a) give the same expression
by transforming the variables].
${\cal S}_1$ indicates the summation taken over
$(s_1,s_2,s_3,s_4)=
(+,+,+,K)$, $(+,+,K,-)$, $(+,K,-,-)$, and $(K,-,-,-)$.
In the same way,
$(+,K,+,K)$, $(+,K, K,-)$,
and $(+,+,-,-)$
for ${\cal S}_2$, and
$(K, K,-,-)$ and $(+,-,-,-)$ for ${\cal S}_3$.
As in the case of the MT term,
in the dirty limit,
the summation over ${\mib k}$ results in the
following expression:
\begin{equation}
  \begin{split}
&   \frac{1}{N^3}\sum_{\mib k}
   (\hat{G}_k^{s_3}v_{k}\hat{G}_{k-q}^{s_4}
   v_{k}\hat{G}_k^{s_1})
    (\hat{G}_{k-q'}^{s_2})
  \simeq 
  \frac{\pi \rho_0\tau^3}{4}
  \int_{FS}v_k^2
  \{
4  (\hat{g}_{\epsilon}^{s_3}\hat{g}_{\epsilon-\omega}^{s_4}
\hat{g}_{\epsilon}^{s_1})
(\hat{g}_{\epsilon-\omega'}^{s_2})\\
&+  [  (\hat{g}_{\epsilon}^{s_3}\hat{g}_{\epsilon-\omega}^{s_4}
    \hat{\tau}_3)
    + (\hat{g}_{\epsilon}^{s_3}\hat{\tau}_3
    \hat{g}_{\epsilon}^{s_1})
    + (\hat{\tau}_3\hat{g}_{\epsilon-\omega}^{s_4}
    \hat{g}_{\epsilon}^{s_1})
+(\hat{\tau}_3)  ](\hat{\tau}_3)
  +[(\hat{g}_{\epsilon}^{s_3}\hat{g}_{\epsilon-\omega}^{s_4}
\hat{g}_{\epsilon}^{s_1})+ (\hat{g}_{\epsilon}^{s_3})
    + (\hat{\tau}_3\hat{g}_{\epsilon-\omega}^{s_4}
    \hat{\tau}_3)
    + (\hat{g}_{\epsilon}^{s_1})]   (\hat{g}_{\epsilon-\omega'}^{s_2})
\}.
  \end{split}
\end{equation}
Then, the result of the self-energy term is given by
Eq. (\ref{eq:condDOS0}).

The real part of the conductivity by the DOS term shown in Fig. 4(b)
is given by Eq. (\ref{eq:DOS0tr}) with
$(1/N^3)\sum_{\mib k}{\rm Tr}[\;\cdot\;]$
replaced by 
\begin{equation}
  \begin{split}
n_i u^2\left(    \frac{1}{N^3}\right)^2\sum_{\mib k,\mib k'}
    {\rm Tr}[\hat{I}^{s_3,b'}_{\epsilon}
      \hat{G}_{k'}^{s_3}v_{k'}
      \hat{G}_{k'-q}^{s_4}
\hat{\tau}_3\hat{I}^{s_4,c'}_{\epsilon-\omega}
      \hat{G}_{k-q}^{s_4}
      v_{k}\hat{G}_k^{s_1}
    \hat{I}^{s_1,a}_{\epsilon}
    \hat{Y}_{i}\hat{I}^{s_2,a'}_{\epsilon-\omega'} 
    \hat{G}_{k-q'}^{s_2}
    \hat{I}^{s_2,c}_{\epsilon-\omega'}\hat{\tau}_3
    \hat{G}_{k'-q'}^{s_2}
    \hat{I}^{s_2,b}_{\epsilon-\omega'}\hat{Y}'_{i}].
      \end{split}
\end{equation}
We calculate this expression with the use of the same approximation
as in Eq. (\ref{eq:MTapprox1}).
The result for the DOS term is given by
Eq. (\ref{eq:condDOS}).

\subsection{Aslamazov--Larkin term}

The real part of the conductivity for the left diagram in Fig. 5
is given by the following expression:
\begin{equation}
  \begin{split}
  {\rm Re}\sigma^{AL1}_{\omega}=&
    \frac{e^2}{2\omega N^3}\sum_{\mib q'}
  \int\frac{d\omega'}{2\pi}\sum_{i,j=0,1,2,3,4}
           {\rm Re}\{
           (C^t_{\omega'}+C^t_{\omega-\omega'})
           \Gamma_i(q')\Gamma_j(q-q')
                 {\cal R}^{(1)}_{i,j}{\cal Q}^{(1)}_{i,j}
                 \\
&+(C^t_{\omega}-C^t_{\omega'})                 
\Gamma^*_i(q')\Gamma_j(q-q')
      {\cal R}^{(2)}_{i,j}{\cal Q}^{(2)}_{i,j}
      +(C^t_{\omega}-C^t_{\omega-\omega'})
\Gamma_i(q')\Gamma^*_j(q-q')
      {\cal R}^{(3)}_{i,j}{\cal Q}^{(3)}_{i,j}
                 \\
&-C^t_{\omega}
\Gamma^*_i(q')\Gamma_j(q-q')
      {\cal R}^{(2)}_{i,j}{\cal Q}^{(4)}_{i,j}
      -C^t_{\omega}
\Gamma_i(q')\Gamma^*_j(q-q')
      {\cal R}^{(3)}_{i,j}{\cal Q}^{(5)}_{i,j}      \}
  \end{split}
  \label{eq:resigmaAL1}
\end{equation}
with
\begin{equation}
  {\cal R}^{(x)}_{i,j}=
  \sum_{{\cal S}_x}
  \int\frac{d\epsilon}{2\pi}\frac{1}{N^3}\sum_{\mib k}
           {\rm Tr}[\hat{G}^{s_1}_{k+q}v_{k}\hat{G}^{s_2}_{k}
\hat{I}^{s_2,a}_{\epsilon}
\hat{Y}'_i
\hat{I}^{s_3,a'}_{\epsilon+\omega'}
\hat{G}^{s_3}_{k+q'}
\hat{I}^{s_3,b}_{\epsilon+\omega'}
\hat{Y}'_j
\hat{I}^{s_1,b'}_{\epsilon+\omega}]
\end{equation}
and
\begin{equation}
  {\cal Q}^{(x)}_{i,j}=
  \sum_{{\cal S}_x}
\int\frac{d\epsilon}{2\pi}\frac{1}{N^3}\sum_{\mib k}
           {\rm Tr}[\hat{G}^{s_2}_k v_{k}\hat{G}^{s_1}_{k+q}
\hat{I}^{s_1,b}_{\epsilon+\omega}
\hat{Y}_j
\hat{I}^{s_3,b'}_{\epsilon+\omega'}
\hat{G}^{s_3}_{k+q'}
\hat{I}^{s_3,a}_{\epsilon+\omega'}
\hat{Y}_i
\hat{I}^{s_2,a'}_{\epsilon}].
\end{equation}
$\sum_{{\cal S}_x}$ indicates that
the summation is taken over $(s_1,s_2,s_3)$ with
\begin{equation}
  (s_1,s_2,s_3)=
  \left\{
  \begin{matrix}
(+,K,+), (+,-,K), (K,-,-) & (x=1),\\
(+,+,K), (+,K,-), (K,-,-) & (x=2),\\
(+,K,+), (K,-,+), (-,-,K) & (x=3),\\
(+,+,K), (K,+,-), (-,K,-) & (x=4),\\
(K,+,+), (-,K,+), (-,-,K) & (x=5).
  \end{matrix}
  \right.
\end{equation}
$(1/N^3)\sum_{\mib k}
           (\hat{G}^{s_2}_k v_{k}\hat{G}^{s_1}_{k+q})
(\hat{G}^{s_3}_{k+q'})$
and
$(1/N^3)\sum_{\mib k}
           (\hat{G}^{s_1}_{k+q}v_{k}\hat{G}^{s_2}_{k})
(\hat{G}^{s_3}_{k+q'})$
are calculated in the same way as in Eq. (\ref{eq:MTapprox1}).
In the same way, the real part of the
conductivity for the right diagram of Fig. 5
(${\rm Re}\sigma^{AL2}_{\omega}$) is calculated, and it is 
given by Eq. (\ref{eq:resigmaAL1})
with ${\cal R}^{(1)}_{i,j}$, ${\cal R}^{(2)}_{i,j}$, and
${\cal R}^{(3)}_{i,j}$
replaced by
${\cal R}^{'(1)}_{i,j}$, ${\cal R}^{'(3)}_{i,j}$,
and ${\cal R}^{'(3)}_{i,j}$,
respectively.
Here,
\begin{equation}
  {\cal R}^{'(x)}_{i,j}=
  \sum_{{\cal S}_x}
  \int\frac{d\epsilon}{2\pi}\frac{1}{N^3}\sum_{\mib k}
           {\rm Tr}[\hat{G}^{s_1}_{k+q}v_{k}\hat{G}^{s_2}_{k}
\hat{I}^{s_2,a}_{\epsilon}
\hat{Y}'_j
\hat{I}^{s_3,a'}_{\epsilon+\omega-\omega'}
\hat{G}^{s_3}_{k+q-q'}
\hat{I}^{s_3,b}_{\epsilon+\omega-\omega'}
\hat{Y}'_i
\hat{I}^{s_1,b'}_{\epsilon+\omega}].
\end{equation}
The summation over $\mib k$
in
$(1/N^3)\sum_{\mib k}
           (\hat{G}^{s_1}_{k+q}v_{k}\hat{G}^{s_2}_{k})
(\hat{G}^{s_3}_{k+q-q'})$
is performed 
in the same way as in the above calculations.

Using the above expressions,
we find that ${\cal R}$, ${\cal Q}$, and ${\cal R}'$ have
the following properties:
${\cal R}^{(x)}_{i,j}={\cal Q}^{(x)}_{i,j}={\cal R}^{'(x)}_{i,j}=0$
for $i=j=0,1,2,3$ and $(i,j)=(0,1),(1,0),(2,3),(3,2)$ with $x=1,2,3$.
${\cal R}^{(1,2,3)}_{i,j}={\cal R}^{'(1,3,2)}_{i,j}$
for $(i,j)=(1,2),(2,1),(1,3),(3,1)$,
and
${\cal R}^{(1,2,3)}_{i,j}=-{\cal R}^{'(1,3,2)}_{i,j}$
for $(i,j)=(0,2),(2,0),(0,3),(3,0)$.
Then, the sum of 
${\rm Re}\sigma^{AL1}_{\omega}$ and
${\rm Re}\sigma^{AL2}_{\omega}$ 
includes
only the terms of
$\Gamma^{(*)}_i(q')\Gamma^{(*)}_j(q-q')$
with $(i,j)=(1,3)$, $(3,1)$, $(1,2)$, $(2,1)$,
$(1,4)$, and $(4,1)$.
The result of the AL term ($\sigma^{AL}_{\omega}
=\sigma^{AL1}_{\omega}+\sigma^{AL2}_{\omega}$)
is given by Eq. (\ref{eq:condAL}).

\subsection{Maximally crossed term}

The real part of the conductivity given in Fig. 6(b) is
expressed as 
\begin{equation}
  \begin{split}
  {\rm Re}\sigma^{MC}_{\omega}=&
  \frac{-e^2}{2\omega N^3}\sum_{\mib q'}\sum_{s=\pm}s
  \int\frac{d\epsilon}{2\pi}(T^h_{\epsilon+\omega}-T^h_{\epsilon})
  \frac{1}{N^3}\sum_{\mib k}v_{\mib k} v_{\mib q'-\mib k}
  \sum_{n=1}^{\infty}
  n_i u^2\left(\frac{n_i u^2}{N^3}\right)^n
  \sum_{\mib k_1,\cdots,\mib k_n}{\rm Tr}[\\
&    \hat{G}^s_{\mib q'-\mib k,\epsilon}
    \hat{G}^+_{\mib q'-\mib k,\epsilon+\omega}
    \hat{\tau}_3\hat{G}^+_{\mib k_n,\epsilon+\omega}
    \cdots\hat{\tau}_3\hat{G}^+_{\mib k_1,\epsilon+\omega}\hat{\tau}_3
    \hat{G}^+_{\mib k,\epsilon+\omega}
    \hat{G}^s_{\mib q'-\mib k,\epsilon}
    \hat{\tau}_3\hat{G}^s_{\mib q'-\mib k_n,\epsilon}
    \cdots\hat{\tau}_3\hat{G}^s_{\mib q'-\mib k_1,\epsilon}\hat{\tau}_3].
  \end{split}
\end{equation}
$(1/N^3)\sum_{\mib k}v_{\mib k} v_{\mib q'-\mib k}
(\hat{G}^s_{\mib q'-\mib k,\epsilon}
    \hat{G}^+_{\mib q'-\mib k,\epsilon+\omega})
(\hat{G}^+_{\mib k,\epsilon+\omega}
    \hat{G}^s_{\mib q'-\mib k,\epsilon})$
    is calculated in the same way as in Eq. (\ref{eq:MT0g4}).
    The calculation similar to Eq. (\ref{eq:MT0diff})
    shows that
\begin{equation}
  \sum_{n=1}^{\infty}
  \left(\frac{n_i u^2}{N^3}\right)^n
  \sum_{\mib k_1,\cdots,\mib k_n}
    (\hat{G}^+_{\mib k_n,\epsilon+\omega}
    \cdots\hat{\tau}_3\hat{G}^+_{\mib k_1,\epsilon+\omega})
    (\hat{G}^s_{\mib q'-\mib k_n,\epsilon}
    \cdots\hat{\tau}_3\hat{G}^s_{\mib q'-\mib k_1,\epsilon})
    \simeq
    \frac{X^{+s}_{\epsilon+\omega,\epsilon}[(\hat{g}^+_{\epsilon+\omega})(\hat{g}^s_{\epsilon})
           +(\hat{\tau}_3)(\hat{\tau}_3)]
}
         {1-2X^{+s}_{\epsilon+\omega,\epsilon}}.
\end{equation}
Then, the real part of the
conductivity for this term is given by Eq. (\ref{eq:condMC}).


\begin{thebibliography}{9}
\bibitem{HStrans}
  The following relation is obtained by
    the Hubbard--Stratonovich transformation: 
    $exp(-i\int_{\cal C}
    d t\sum_{\mib k,\mib k',\mib q,\sigma,\sigma'}
  v_{\mib q}^C\bar{\psi}_{\mib k,\sigma,t}\psi_{\mib k+\mib q,\sigma,t}
  \bar{\psi}_{\mib k',\sigma',t}\psi_{\mib k'-\mib q,\sigma',t}/2N^3)
  =\int{\cal D}[\varphi]exp(-i\int_{\cal C} d t\sum_{\mib q}[\varphi_{\mib q,t}
    \varphi_{-\mib q,t}e^2/2v^C_{\mib q}
    +i e\varphi_{\mib q,t}\sum_{\mib k,\sigma}\bar{\psi}_{\mib k,\sigma,t}
    \psi_{\mib k-\mib q,\sigma,t}/\sqrt{N^3}])$.
  Using $\int_{\cal C} d t\varphi_{\mib q,t}\varphi_{-\mib q,t}=
  \int_{-\infty}^{\infty} d t(\varphi^+_{\mib q,t}\varphi^+_{-\mib q,t}
  -\varphi^-_{\mib q,t}\varphi^-_{-\mib q,t})=
  \int_{-\infty}^{\infty} d t(\varphi^{cl}_{\mib q,t}\varphi^{qu}_{-\mib q,t}
  +\varphi^{qu}_{\mib q,t}\varphi^{cl}_{-\mib q,t})$,
  we obtain the second term of Eq. (\ref{eq:actionSd}).
  Here, the superscript $+$ $(-)$ in $\varphi_{\mib q,t}^{\pm}$
  indicates that the field $\varphi_{\mib q,t}$ resides
  on the forward (backward) branch.
  A calculation of the term
  $\int_{\cal C} d t \varphi_{\mib q,t}\bar{\psi}_{\mib k,\sigma,t}
  \psi_{\mib k-\mib q,\sigma,t}$ is performed in a similar way.~\cite{kamenev}
\bibitem{memo1}
    We use this one-particle Green's function to perturbatively
    calculate the weak localization correction to the linear absorption.
    In this calculation, the weak localization effect is not
    included in a self-consistent way in the one-particle state
    of electrons. This method of calculation is the same as that
    previously
    used in the case of the normal state,~\cite{AASSC,AAJETP,fukuyama,AALee,ovchinnikov,maekawa,takagi,schmidt,aslamasovV,federici,petkovic,altshuler}
    and is considered to be valid also in the superconducting state.
    This is because the excitation energy of
        the superconducting phase fluctuation is
        pushed up to the plasma frequency, and
        the effect of fluctuation becomes small
        in a three-dimensional system.
\bibitem{memo2}
    In the superconducting state,
    the weak localization effect lowers $T_c$
    and reduces the superconducting gap $\Delta$.~\cite{smith95}
    In this perturbative calculation,
    the value of $\Delta$ itself does not affect the spectrum
    qualitatively
        ($\Delta$ at $T=0$
    is set to be the unit of energy).
    The validity of this method of calculation is related to
    the behavior of the spectrum around the gap edge.
    For example, when the effect of paramagnetic impurities
    on the spectrum is considered perturbatively 
    (Sect. 4 in Ref. 25),
    the perturbation
    term diverges at the gap edge.
    In this case, the spectrum should be calculated consistently
    with the values of $\Delta$. On the other hand,
    in the weak localization correction, the perturbation term for
    the spectrum is finite, and then the behavior of the spectrum
    can be discussed independently of the size of the superconducting gap.
\bibitem{memo3}
    In a three-dimensional system,
      the higher-order terms 
      are proportional to
      $1/(k_Fl)^{2n}$ ($n$ is 
      the number of the irreducible four-point interaction
      vertices included in the conductivity) for all the cases of
      ``MT0'', ``MT'', ``DOS0'', ``DOS'', ``AL'', and ``MC''
      in the dirty limit $\Delta\tau\ll 1$ (the mean free path
      is shorter than the coherence length).
      Our calculation is valid in the case of
      $(k_Fl)^2\gg 1$.
      On the other hand, in the case of a (quasi-)two-dimensional
      system,
      the $n$-th order term in 
      ${\rm Re}\sigma_{\omega}$ is proportional to
      $1/(k_Fl)^n$. 
      Then, in a thin film, the higher-order terms are more effective
      than in a three-dimensional case, and
      the approximation used in this study is valid
      in the latter case for a wide range of values of $k_Fl$.
\bibitem{memo4}
  The subscripts $i=0,1,2,3$ in $\chi_i$ and $\Gamma_i$
    are related
  to those of Pauli matrices ($\hat{\tau}_i$) in the Nambu space.
  For example, $\hat{\tau}_{3}$ and $\hat{\tau}_2$
  indicate the density fluctuation and
  the superconducting phase fluctuation,
  respectively, as noted in Ref. 25.
  $\Gamma_4(q)$ originates from the mixing term $\chi'$, which
  couples the above two fluctuations.
\end{thebibliography}
\end{document}